\documentclass[12pt,letterpaper]{article}
\usepackage[utf8]{inputenc}
\DeclareUnicodeCharacter{03B8}{\circ}
\usepackage{titling}
\usepackage{scalerel}
\usepackage{textcomp}
\usepackage{gensymb}
\usepackage{graphicx}
\usepackage{scalerel}
\usepackage{amsmath}
\usepackage{booktabs}
\usepackage{tabu}
\usepackage{graphicx}
\usepackage{caption}
\usepackage{subcaption}
\usepackage{cite}
\usepackage{secdot}
\usepackage[margin=0.8in]{geometry}
\usepackage{amssymb}
\usepackage{algorithm2e}
\usepackage[affil-it]{authblk}
\usepackage{abstract}

\usepackage[nodisplayskipstretch]{setspace}
\providecommand{\keywords}[1]
{
  \small	
  \textbf{Keywords:} #1
}
\usepackage{url}
\usepackage[colorlinks = true,
            linkcolor = blue,
            urlcolor  = blue,
            citecolor = blue,
            anchorcolor = blue]{hyperref}\usepackage{ragged2e}
\begin{document} 
\title{\large{\textbf{
  Connecting Neutrino Masses, Dark Matter and Leptogenesis \\from \boldmath$\Delta(54)$ Flavor with Triple Inverse Seesaw
 \vspace{-0.45em}}}}

\author{
H. Bora$^{1} \footnote{
\textit{Corresponding author: }
\href{mailto:hrishi@tezu.ernet.in}{hrishibora1997@gmail.com}
}$, 
N. Bharali$^2$,
R. Sarkar$^3$,
S.K. Jha$^4$, 
A. Baruah$^5$ 
and Ng.K. Francis$^6$
}

\setlength{\affilsep}{1em}

\affil{$^1$\normalsize  Department of Physics, University of Science and Technology, Baridua-793101, India }

\affil{\vspace{-1.05em}$^2$\normalsize{Centre of Plasma Physics, Institute for Plasma Research, Sonapur-782402, India}}

\affil{\vspace{-1.05em}$^3$\normalsize{Department of Physics, Rajiv Gandhi Institute of Petroleum Technology, Sivasagar-785697,India}}

\affil{\vspace{-1.05em}$^4$\normalsize{Department of Physics, Indian Institute of Technology Guwahati, Guwahati-781039, India}}

\affil{\vspace{-1.05em}$^5$\normalsize{Department of Physics, Banaras Hindu University, Varanasi-221005, India}}

\affil{\vspace{-1.05em}$^6$\normalsize{School of Physical Sciences, Jawaharlal Nehru University, New Delhi-110067, India}}

\date{\vspace{-1ex}}
\vspace{-1cm}
\maketitle
\vspace{-1.2cm}
\begin{center}
\textbf{\large{Abstract}}\\
\justify

In this present study, an extended $\Delta(54)$ flavor symmetry model incorporating  two standard model Higgs doublets is investigated. This model generates neutrino masses through the triple inverse seesaw mechanism. It predicts deviations from tribimaximal mixing, yielding a nonzero reactor angle ($\theta_{13}$) and an atmospheric mixing angle ($\theta_{23}$) in the upper octant. In addition, the CP-violating phase and the Jarlskog invariant are found to be consistent with current neutrino oscillation data. Our study also includes the dark matter sector by evaluating the relic abundance and active neutrino–dark matter mixing under relevant cosmological constraints. Furthermore, baryogenesis is achieved through resonant leptogenesis at the TeV scale including flavor effects. We obtain the observed baryon asymmetry, $\eta_B \approx 6 \times 10^{-10}$, for right-handed neutrino mass $M_1 = 10$ TeV and mass splitting $d \approx 10^{-8}$.
\end{center}

\keywords{Flavor~symmetry; Inverse seesaw; Neutrino~mixing; Dark~matter; Resonant~leptogenesis}

\newpage
\section{Introduction}
\label{sec:intro}

Neutrino oscillations limits the Standard Model (SM), which predicts massless neutrinos. In contrast, it demonstrates that at least two neutrinos possess distinct, non-degenerate masses. While experiments and theoretical models has reported neutrino masses, with the type-I seesaw mechanism  as early as in 1977. Subsequently, numerous neutrino mass models and generation mechanisms have been developed \cite{chulia2021inverse}. These frameworks primarily aim to generate non-zero neutrino masses \cite{DayaBay:2012fng, RENO:2012mkc ,MINOS:2011amj, DoubleChooz:2011ymz, T2K:2011ypd } and  explain the smallness of neutrino masses relative to other fermions. Typically, this is achieved through seesaw mechanisms \cite{minkowski1977mu, yanagida1980horizontal, mohapatra1980neutrino}, loop mechanisms \cite{zee1980theory, cheng1980neutrino, zee1986quantum }, naturalness mechanisms \cite{mohapatra1986neutrino, Ma2001}, hybrid mechanisms \cite{Aoki2009} etc. 

Although this work concentrates exclusively on Majorana neutrino models, it is essential to acknowledge that the fundamental nature of neutrinos remains unresolved. Consequently, both Dirac and Majorana neutrino possibilities must be considered in a comprehensive examination of neutrino properties. This study undertakes an in-depth examination of the triple inverse seesaw mechanism, a prominent naturalness framework. Focusing on Majorana neutrinos, we elucidate the diverse possibilities inherent to this mechanism, with particular emphasis on its Majorana realization. Notably, the majority of models presented herein are, to our knowledge, novel constructions introduced for the first time in this work.

At low energies, the generation of Majorana neutrino masses can be described by an effective operator \cite{chulia2018seesaw, anamiati2018high}, 
\begin{equation}
    \mathcal{L}_M = \frac{C_M}{\Lambda^{m+n-1}} \bar{L^c} L \Phi^{m} \sigma^{n} - h.c., 
\end{equation}
where $C_M$ represents the effective coupling constant matrix, $L$ denotes the standard model lepton doublet and $\Lambda$ signifies the energy scale associated with new degrees of freedom that generate this effective operator.

The inverse seesaw mechanism offers a viable framework for generating neutrino masses, with mediator masses potentially accessible at the electroweak scale. This approach is distinguished by a small mass parameter $\mu$, which adheres to the hierarchical structure 
\begin{equation}
    \mu << v << \Lambda
\end{equation}
where $v$ represents the Higgs vacuum expectation value  and $\Lambda$ denotes the scale of neutrino mass generation. The $\mu$ parameter acts as a suppressor of neutrino masses, scaling as $m_v \propto  \mu$, thereby enabling the reproduction of observed neutrino masses and mixing angles through large Yukawa couplings and relatively light seesaw mediators. This typically yields a more diverse phenomenological landscape compared to traditional high energy seesaw scenarios.

The neutrino phenomenology of the inverse seesaw is dependant upon the presence of an additional singlet fermion \cite{kakoti2025kev, bora2024relic}. According to this formalism, the lightest neutrino mass matrix is $M_{\nu}\approx M_d (M_T)^{-1}\mu M^{-1} {M_d}^T$ where $M$ denotes the lepton number conserving interaction between right-handed and sterile fermions, $M_d$ is the Dirac mass term and $\mu$ is the Majorana mass term.
Three more right-handed neutrinos and one gauge singlet chiral fermion field $S$
as a sterile neutrino are added to the standard model particles in the Minimal Extended see-saw (MES), which is an extension of the canonical type-I see-saw.
The $4\times 4$ active-sterile neutrino mass matrix in this formalism is given by \cite{zhang2012light,bora2024relic}
\begin{equation}   \label{eq:1} 
M_{\nu}^{4\times4}= - \begin{pmatrix}
     M_D M^{-1}_R M^T_D &  M_D M^{-1}_R M^T_S \\
     M_S (M^{-1}_R)^T M^T_D &   M_S M^{-1}_R M^T_S \\
    \end{pmatrix} 
\end{equation}
where $M_D$, $M_R$ and $M_S$ are the Dirac, Majorana and Sterile neutrino mass matrices. The  Sterile neutrino mass is given as $\label{eq3}
   m_s \simeq -M_S M^{-1}_R M^T_S$.\\
   
Neutrinos with sub-eV can be produced from $M_D$ at the electroweak scale, $M_R$ at the TeV scale, and $M_S$ at the keV scale.
There are other important lines of evidence for dark matter (DM), such as Fritz Zwicky's 1933 observations of galaxy clusters, gravitational lensing (which Zwicky 1937 proposed could allow galaxy clusters to act as gravitational lenses), galaxy rotation curves in 1970, cosmic microwave background, and the most recent cosmology data provided by the Planck satellite.
It is known that dark matter (DM) makes up around 27\% of the universe, which is almost five times more than baryonic matter, based on the most current data from the Planck spacecraft. According to reports, the current dark matter abundance is given as\cite{taoso2008dark,bora2024relic}
\begin{equation*}
    \Omega_{DM} h^2 = 0.1187 \pm 0.0017
\end{equation*}\\
The physics community has faced enormous challenges in its search for potential dark matter candidates with new mechanisms beyond the standard model. The important criteria that a particle must have in order to be taken into consideration as a viable DM candidate. All of the SM particles are not eligible to be DM candidates due to these restrictions. The particle physics community became motivated to investigate several BSM frameworks that may provide accurate DM phenomenology and can be evaluated in many experiments\cite{bora2024relic}.

The extension of the standard model with ISS mechanism can explain the observed Baryon Asymmetry of the Universe (BAU) through leptogenesis \cite{fukugita1986barygenesis}.  In leptogenesis, the asymmetry in leptons, obtained from the CP-violating decay of heavy right-handed neutrinos, is transformed into an asymmetry in baryons through sphaleron processes \cite{kuzmin1985anomalous}. According to Ref. \cite{davidson2002lower},  a mass scale of around $\mathcal{O}(10^9)$ for the right-handed neutrino is necessary to explain the observed BAU. However, this requirement can be reduced if the masses of right-handed neutrinos are nearly the same. In such cases, the effects that violate CP symmetry become significantly amplified, and with relatively low masses (TeV scale), sufficient asymmetry in leptons can be generated to account for the Baryon Asymmetry of the Universe (BAU). This condition is termed resonant leptogenesis. It is important to mention that recent research, utilizing the $SU(5) \times \mathcal{T}_{13}$ model, has shown the possibility of resonant leptogenesis at the GeV–TeV scale within the type-I seesaw model, considering active sterile mixing within the sensitivity range of DUNE. Additionally, considerable attention has been given to investigating the origin of neutrino flavor mixing. Among the available explanations, Tri-bimaximal mixing (TBM) appears to be the most probable\cite{bora2026neutrino}. However, experimental results from Daya Bay, RENO, and Double Chooz suggest that TBM needs to be adjusted to incorporate a non-zero value for $\theta_{13}$ \cite{DayaBay:2012fng, DoubleChooz:2011ymz}. The $\Delta$(54) symmetry can
manifest itself in heterotic string models on factorizable orbifolds, such as the $T^2/Z_3$ orbifold \cite{ ishimori2009lepton, singh2022trimaximal}. In these string models, only singlets and triplets are observed as fundamental modes, while doublets are absent as fundamental modes. However, doublets have the potential to become fundamental modes in intersecting D-brane models. By employing resonant leptogenesis, the resulting mass matrix can potentially explain the Baryon Asymmetry of the Universe (BAU) concurrently. To achieve successful resonant leptogenesis, we introduce a higher-order term. We specifically choose the Majorana mass matrix for right-handed neutrinos, $M_{R}$, so that these neutrinos have degenerate masses at the dimension five-level. In essence, our work expands upon the model proposed in \cite{bora2023neutrino}, making it suitable for investigating resonant leptogenesis in scenarios involving the minimum seesaw model\cite{bora2026neutrino}.

The authors of a prior study proposed the use of the Inverse Seesaw mechanism in conjunction with the $\Delta(54)$ flavor model for Dirac neutrinos\cite{bora2023neutrino}. This study provides a demonstration of the Triple Inverse Seesaw mechanism for Majorana neutrinos, using two Standard Model Higgs \cite{bora2026neutrino}. We introduced a Vector-Like (VL) and these additional components produces a Majorana mass term after the symmetry breaking. 

The structural outline of our paper is as follows: The framework of the model, including the fields and their symmetrical transformation characteristics, is outlined in Section \ref{frame}. The neutrino phenomenological findings are numerically analyzed \cite{bora2026neutrino}and examined in Section  \ref{num}.  In Section \ref{DM}, we summarize the  dark matter sector.  The baryon asymmetry along with the framework for resonant leptogenesis is presented in Section \ref{Res}. We
conclude with our final remarks in Section  \ref{conc}.

\section{Framework of the Model}
\label{frame}
To realize the Majorana triple inverse seesaw mechanism, extending the Standard Model fermion sector is crucial. We achieve this by augmenting the $\Delta(54)$ flavor symmetry model, incorporating two Standard Model Higgs fields ($H$ and $H^\prime$) and distinct flavons. In this study, we have presented two VL fermions denoted as $N_i$ and $S_i$, which possess the characteristic of being gauge singlets inside the framework of the Standard Model.  The subscript $i$ can be $R$ or $L$ associated with right-handedness and left-handedness respectively.

Our proposed model builds upon the $\Delta(54)$ framework, introducing supplementary flavons to accommodate deviations from the ideal Tri-Bimaximal (TBM) neutrino mixing paradigm\cite{2010}. We put extra symmetry $Z_2\otimes Z_3 \otimes Z_4$ to avoid undesirable terms. Table \ref{tab:2} provides details regarding the composition of the particles and corresponding charge assignment in accordance with the symmetry group. The triplet representation of $\Delta(54)$ is used to assign the left-handed leptons doublets and the right-handed charged lepton. The representations of $\Delta(54)$ symmetry are real that guarantees the construction of the effective Lagrangian.

\begin{table}[ht]
    \centering
    \scalebox{0.8}{
 {\begin{tabular}{c c c c c  c c c c c c c c c c c c c}
     \hline
       \textrm{Field}  &  L & $l $ & $H$ & $H^{\prime}$
  & $N_R$ & $N_L$  & $S_R$ & $S_L$ & $\chi$ & $\chi^\prime$ & $\zeta$ & $\zeta^\prime$ & $\xi$ & $\xi^{\prime}$ & $\phi$ & $\phi^\prime$ & $\phi^{\prime\prime}$\\
     \hline
     \textrm{$\Delta(54)$}  &  $3_{1(1)}$ &  $3_{2(2)} $ & $1_{1}$ & $1_{2}$ & $3_{1(1)}$ & $3_{2(1)}$  & $3_{2(2)}$ & $3_{1(2)}$ & $1_2$ & $2_1$ & $1_{1}$ & $1_{1}$ & $3_{2(1)}$ & $3_{1(1)}$ & $3_{2(2)}$ & $3_{2(1)}$ & $3_{1(2)}$\\
     \textrm{Z}$_2$  &  1 & -1 & 1 & 1 & -1 & 1 & 1 & 1 & -1 & -1 & -1 & 1 & -1 & -1 & 1 & -1 & 1\\
    \textrm{Z}$_3$  &  $\omega$ & $\omega$ & 1 & 1  & 1 & $\omega$  & $\omega$  & $\omega$  & 1 & 1& $\omega$ & 1 & $\omega$ & $\omega$ & 1 & $\omega^2$ & 1\\
\textrm{Z}$_4$  &  1 & -1 & 1 & 1  & 1 & -1  & 1 & 1 & -1 &-1& 1 & -1 & 1 & 1 & 1 &-1& 1\\
     \hline
    \end{tabular}}}
    \caption{Full particle content of our model}
    \label{tab:1}
    \end{table}

The Lagrangian is as follows \footnote{Considering terms upto dimension-5.}:
\begin{align*}
  \mathcal{L} = & \frac{y_1}{\Lambda} ( l \Bar{L} ) \chi H + \frac{y_2}{\Lambda} ( l \Bar{L} ) \chi^\prime H   + \frac{\Bar{L} \Tilde{H^{\prime}} N_{R}}{\Lambda}y_{\scaleto{\xi}{5pt}} \xi + \frac{\Bar{L} \Tilde{H} N_{R}}{\Lambda} y_{\scaleto{S}{3pt}}\xi^{\prime}     \\                 
 & + y_{\scaleto{NS}{3pt}}  \Bar{S_L} N_R \zeta  + y_{\scaleto{NS}{3pt}}^{\prime} S_{R}\Bar{N_{L}} \zeta^\prime  + y_{\scaleto{S}{3pt}}  S_{L}\Bar{S_{R}} \phi +  y_{\scaleto{N}{3pt}} N_{L}\Bar{N_{R}} \phi^{\prime}   +  y_{\scaleto{\mu}{4pt}} S^c_R \Bar{S_R}\phi^{\prime\prime}  + y_{\scaleto{\mu1}{5pt}}^{\prime} S_L\Bar{S_L}\phi^{\prime\prime} +  y_{\scaleto{\mu2}{5pt}}^{\prime} S_L\Bar{S_L}\phi^{\prime\prime}
 \label{eq1}
\end{align*}

In this context, the vacuum expectation values are considered naturally as,
\begin{align*}
\langle \chi \rangle& =(v_{\chi})&
\langle \chi^\prime \rangle& =(v_{{\chi}^\prime},v_{{\chi}^\prime})&
\langle \xi \rangle& =(v_{\xi}, v_{\xi}, v_{\xi})&
\langle \xi^\prime \rangle& =(v^{\prime}_{\xi} ,v^{\prime}_{\xi},v^{\prime}_{\xi})&
\\
\langle \zeta \rangle& =(v_{\zeta})&
\langle \zeta^\prime \rangle& =(v_{\zeta}^\prime) &  
\langle \phi \rangle& =(v_{\phi},v_{\phi},v_{\phi})&
\langle \phi^\prime \rangle& =(v^\prime_{\phi},v^\prime_{\phi},v^\prime_{\phi}) & \langle \phi^{\prime\prime} \rangle& =(v^{\prime\prime}_{\phi},v^{\prime\prime}_{\phi},v^{\prime\prime}_{\phi})
\end{align*}

The charged lepton mass matrix is given as \cite{ishimori2009lepton} 
\begin{align*}
    M_l= \frac{y_{1} v}  {\Lambda}
    \begin{pmatrix}
    v_{\chi} & 0 & 0\\
    0 & v_{\chi}  & 0\\
    0 & 0 &  v_{\chi} 
    \end{pmatrix}  +  
    \frac{y_{2} v}  {\Lambda}
    \begin{pmatrix}
    -\omega v_{\chi^\prime} + v_{\chi^\prime} & 0 & 0\\
    0 & -\omega^2 v_{\chi^\prime} + \omega^2 v_{\chi^\prime}  & 0\\
    0 & 0 &   -v_{\chi^\prime} + \omega v_{\chi^\prime}
    \end{pmatrix} 
\end{align*}

where, $y_1$ and $y_2$ are coupling constants and  $v \simeq 55$ GeV.

\subsection{Effective neutrino mass matrix}

The neutrino sector mass matrices are derived from the aforementioned Lagrangian, following the simultaneous breaking of $\Delta(54)$ and electroweak symmetries. Notably, the ISS mechanism facilitates TeV-scale neutrinos, enabling heavy neutrinos to remain remarkably light ($\sim$TeV) while permitting sizable Dirac masses, comparable to those of charged leptons. This framework agrees the coexistence of light neutrino masses ($\sim$sub-eV) with appreciable Dirac Yukawa couplings.

 \begin{align}
&M_{NS} =   y_{\scaleto{NS}{3pt}} \begin{pmatrix}
     v_{\zeta} & 0 & 0\\
    0 &  v_{\zeta} & 0\\
    0 & 0 &  v_{\zeta}
    \end{pmatrix};
   & M^{\prime}_{NS}=y^{\prime}_{\scaleto{NS}{3pt}} \begin{pmatrix}
     v^{\prime}_{\zeta} & 0 & 0\\
    0 &  v^{\prime}_{\zeta} & 0\\
    0 & 0 &  v^{\prime}_{\zeta}
    \end{pmatrix}   \\
&M_{S}=y_{\scaleto{S}{3pt}} \begin{pmatrix}
     v_{\phi} & 0 & 0\\
    0 & v_{\phi}  & 0\\
    0 & 0 &  v_{\phi} 
    \end{pmatrix};
&M_{N}=   y_{\scaleto{N}{3pt}} \begin{pmatrix}
     v^{\prime}_{\phi} & 0 & 0\\
    0 & v^{\prime}_{\phi}  & 0\\
    0 & 0 &  v^{\prime}_{\phi} 
    \end{pmatrix}    \\
&M_\mu=  y_\mu\begin{pmatrix}
     v^{\prime\prime}_{\phi} & 0 & 0\\
    0 &  v^{\prime\prime}_{\phi} & 0\\
    0 & 0 &   v^{\prime\prime}_{\phi}
    \end{pmatrix} ;
 &M_\mu^{\prime}=  y^{\prime}_\mu
 \begin{pmatrix}
    v^{\prime\prime}_{\phi} & 0 & 0\\
    0 & v^{\prime\prime}_{\phi} & 0\\
    0 & 0 &  v^{\prime\prime}_{\phi}
    \end{pmatrix}
    \end{align}
 \begin{equation}
 M_{\nu N}= \frac{v}{\Lambda}        \begin{pmatrix} 
    y_{\scaleto{\xi}{6pt}} v_{\scaleto{\xi}{6pt}} & y_{s}v^{\prime}_{\xi}  & y_{s}v^{\prime}_{\xi} \\
   y_{s}v^{\prime}_{\xi} &   y_{\scaleto{\xi}{6pt}} v_{\scaleto{\xi}{6pt}} & y_{s}v^{\prime}_{\xi} \\
    y_{s}v^{\prime}_{\xi}  & y_{s}v^{\prime}_{\xi} &   y_{\scaleto{\xi}{6pt}} v_{\scaleto{\xi}{6pt}}
    \end{pmatrix}
    \label{eq6}
   \end{equation}\\
Using this hierarchy we obtain the effective neutrino mass matrix as \\  
\begin{equation} \label{eq8}
m_\nu = M^2_{\nu N} \frac{M^{\prime 2}_{NS} M^{\prime}_{\mu}}{M^2_{N}M^2_{S}-2M_{S}M_{N} M^{\prime}_{NS} M_{NS} +  M^{\prime 2}_{NS} M^2_{NS} - M^2_{N} M^2_{\mu} M^{\prime 
2}_{\mu}} 
\approx 
M^2_{\nu N}\frac{M^{\prime 2}_{NS} M^{\prime}_\mu}{M^2_{N}M^2_{S}}
\end{equation}

\begin{equation}
  m_\nu=  \lambda
    \begin{pmatrix}
    \frac{s}{M^2} (2a^2 + b^2)   &    \frac{s}{M^2}(a^2 + 2ab)   &       \frac{s}{M^2}(a^2 + 2ab)\\
      \frac{s}{M^2} (a^2 + 2ab)    &  \frac{s}{M^2}( 2a^2 + b^2 )  &       \frac{s}{M^2}( a^2 + 2ab) \\
   \frac{s}{M^2}(a^2 + 2ab) &   \frac{s}{M^2}(a^2 + 2ab)   &    \frac{s}{M^2}(2a^2 + b^2) 
    \end{pmatrix}\\   
\end{equation}
where $\lambda = \frac{ v^{ 2}}{\Lambda^2} $, $a=y_{\xi} v_{\xi}$ and $b=y_{s} v^\prime_{\scaleto{\xi}{6pt}}$, $M= y_{\scaleto{N}{3pt}}v^{\prime}_{\phi}$ and $s=  \frac{ v^{\prime 2}_{\zeta} y^{\prime 2}_{\scaleto{NS}{3pt}} y^{\prime}_{\mu}v^{\prime\prime}_\phi} { v^2_{\phi}   y^{\prime 2}_{\scaleto{S}{3pt}}}$ 
 
In direct analogy with the standard inverse seesaw, contributions provided from $M_{NS}$ and $M_\mu$ can be safely neglected. Note that here the suppression mechanism is enhanced by three matrices, $M^{\prime 2}_{NS}$ and $M^{\prime}_{\mu}$.
Therefore, we would call this mechanism triple Majorana inverse seesaw \cite{chulia2021inverse}.\\
In the ISS framework, the effective neutrino mass matrix of Eq.\ref{eq8} can be written as
\begin{align} 
m_\nu = & M_{\nu N}(M_{N})^{-1}M_{mid}M_{N}^{-1}M^T_{\nu N}  \quad \text{with} \quad M_{mid} = M_{NS}^{\prime 2}{
  (M^2_S)}^{-1}M^\prime_{\mu} 
\label{eq11}
\end{align}
Here, $M_{\nu N}$ denotes the Dirac mass matrix connecting the active neutrinos and the heavy neutral fermions, while $M_N$ represents the corresponding heavy Majorana mass matrix. The matrix $M_{mid}$  effectively acts as a sterile neutrino mass matrix, generated after integrating out the heavier singlet states. In scenarios where the lightest sterile state is sufficiently stable and feebly mixed with the active neutrinos, it can naturally serve as a viable dark matter candidate.
The final neutrino mixing matrix for the active-sterile mixing takes 
form as,
\begin{equation} \label{eq13}
V \simeq  \begin{pmatrix} (1-\frac{1}{2} R R^\dagger)U & R\\
    -R^\dagger U &  1-\frac{1}{2} R R^\dagger
\end{pmatrix}   
\end{equation}
where $R =   M_D M^{-1}_R M^T_S ( M_S M^{-1}_R M^T_S)$ is a  matrix representing the strength of active
sterile mixing and $U$ is the leptonic mass matrix for active neutrinos\cite{ bora2024relic}. 

\noindent The neutrino mass matrix $m_\nu$ can be diagonalized by the PMNS matrix $U$ as
\begin{equation}
    U^\dagger m^{(i)}_\nu U^* = \textrm{diag(}m_1, m_2, m_3 \textrm{)}
\end{equation}
 We can analytically calculate $U$ using the relation $U^\dagger h U = \textrm{diag(}m_1^2, m_2^2, m_3^2 \textrm{)}$, where $h = m_\nu m^{\dagger}_\nu$. We followed the framework of Adhikary et al. \cite{adhikary2013masses} for  calculating oscillation parameters of a generalized neutrino mass
matrix. The row-wise elements of $U$ are given in terms of the elements
of the $h$ and its eigenvalues $m^2_i$\cite{bora2023neutrino,bora2024relic}.
\begin{equation*}
       U_{1i} = \frac{(h_{22} - m^2_i)h_{13} - h_{12}h_{23}}{N_i}; \quad  U_{2i} = \frac{(h_{11} - m^2_i)h_{23} -h^{*}_{12}h_{13}}{N_i}; \quad
   U_{3i} = \frac{\lvert h_{12} \rvert^2 - (h_{11}-m^2_i)(h_{22}- m^2_i) }{N_i} 
\end{equation*}

\section{Neutrino Masses and Mixing} 
\label{num}
This section presents a numerical examination of the impact of flavon induced corrections on TBM mixing with a focus on the normal hierarchical neutrino mass spectrum. The neutrino mass matrix $m_{\nu}$ can be diagonalized by the PMNS matrix $U$ as \cite{lei2020minimally},
\begin{equation}
    \label{eq:12}
    U^\dagger m_\nu U^* = \textrm{diag(}m_1, m_2, m_3 \textrm{)}
\end{equation}
We can compute the mixing matrix $U$ numerically using the equation $U^\dagger M_\nu U = \textrm{diag}(m_1^2, m_2^2, m_3^2)$, where $M_\nu = m_\nu m^\dagger_\nu$. Subsequently, the neutrino oscillation parameters $\theta_{12}$, $\theta_{13}$ and $\theta_{23}$ from $U$ as \cite{lei2020minimally},
\begin{equation}
    \label{eq:13}
    sin^2{\theta_{12}} = \frac{\lvert U_{12}\rvert ^2}{1 - \lvert U_{13}\rvert ^2}, ~~~~~~ sin^2 {\theta_{13}} = \lvert U_{13}\rvert ^2, ~~~~~~ sin^2{\theta_{23}} = \frac{\lvert U_{23}\rvert ^2}{1 - \lvert U_{13}\rvert ^2}
\end{equation}
and $\delta$ may be given by
\begin{equation}
    \label{eq:14}
    \delta = \textrm{sin}^{-1}\left(\frac{8 \, \textrm{Im(}h_{12}h_{23}h_{31}\textrm{)}}{P}\right)
\end{equation}
with 
\begin{equation}
    \label{eq:15}
     P = (m_2^2-m_1^2)(m_3^2-m_2^2)(m_3^2-m_1^2)\sin 2\theta_{12} \sin 2\theta_{23} \sin 2\theta_{13} \cos \theta_{13}
\end{equation}

\begin{table}[t]
\centering
 \scalebox{1}{
  \begin{tabular}{ l  c  r }
    \hline
    Parameters & NH (3$\sigma$) & IH (3$\sigma$) \\  \hline
    $\Delta{m}^{2}_{21}[{10}^{-5}eV^{2}]$ & $7.236 \rightarrow 7.823$ & $7.236 \rightarrow 7.822$ \\ 
    $\Delta{m}^{2}_{31}[{10}^{-3}eV^{2}]$ & $2.454 \rightarrow 2.592$ & $-2.569 \rightarrow -2.430$\\ 
    $\sin^{2}\theta_{12}$ & $0.2893 \rightarrow 0.3295$ & $0.2893 \rightarrow 0.3295$ \\ 
     $\sin^{2}\theta_{13}$ & $0.0207 \rightarrow 0.0242$ & $0.0209 \rightarrow 0.0243$ \\ 
    $\sin^{2}\theta_{23}$ & $0.432 \rightarrow 0.587$ & $0.437 \rightarrow 0.590$ \\
    $\delta^{\circ}_{CP}$ & $114 \rightarrow 405$ & $202 \rightarrow 347$ \\ \hline
  \end{tabular}}
  \caption{ The neutrino oscillation parameters from NuFIT 6.1 (2025) \cite{nufit61}}
    \label{tab:2}
\end{table}

The 3$\sigma$ ranges of neutrino oscillation parameters from NuFIT 6.0 \cite{esteban2024nufit}. We adjusted the modified $\Delta(54)$ model to suit the experimental data by minimizing the ensuing $\chi^2$ function in order to evaluate how the neutrino mixing parameters contrast with the most current experimental data:

\begin{equation}
	\label{eq:16}
	\chi^2 = \sum_{i}\left(\frac{\lambda_i^{model} - \lambda_i^{expt}}{\Delta \lambda_i}\right)^2,
\end{equation}

where $\lambda_i^{model}$ is the $i^{th}$ observable predicted by the model, $\lambda_i^{expt}$ stands for  $i^{th}$ experimental best-fit value and $\Delta \lambda_i$ is the 1$\sigma$ range of the observable.

\begin{figure}[t]
    \centering
    \includegraphics[width=1\linewidth]{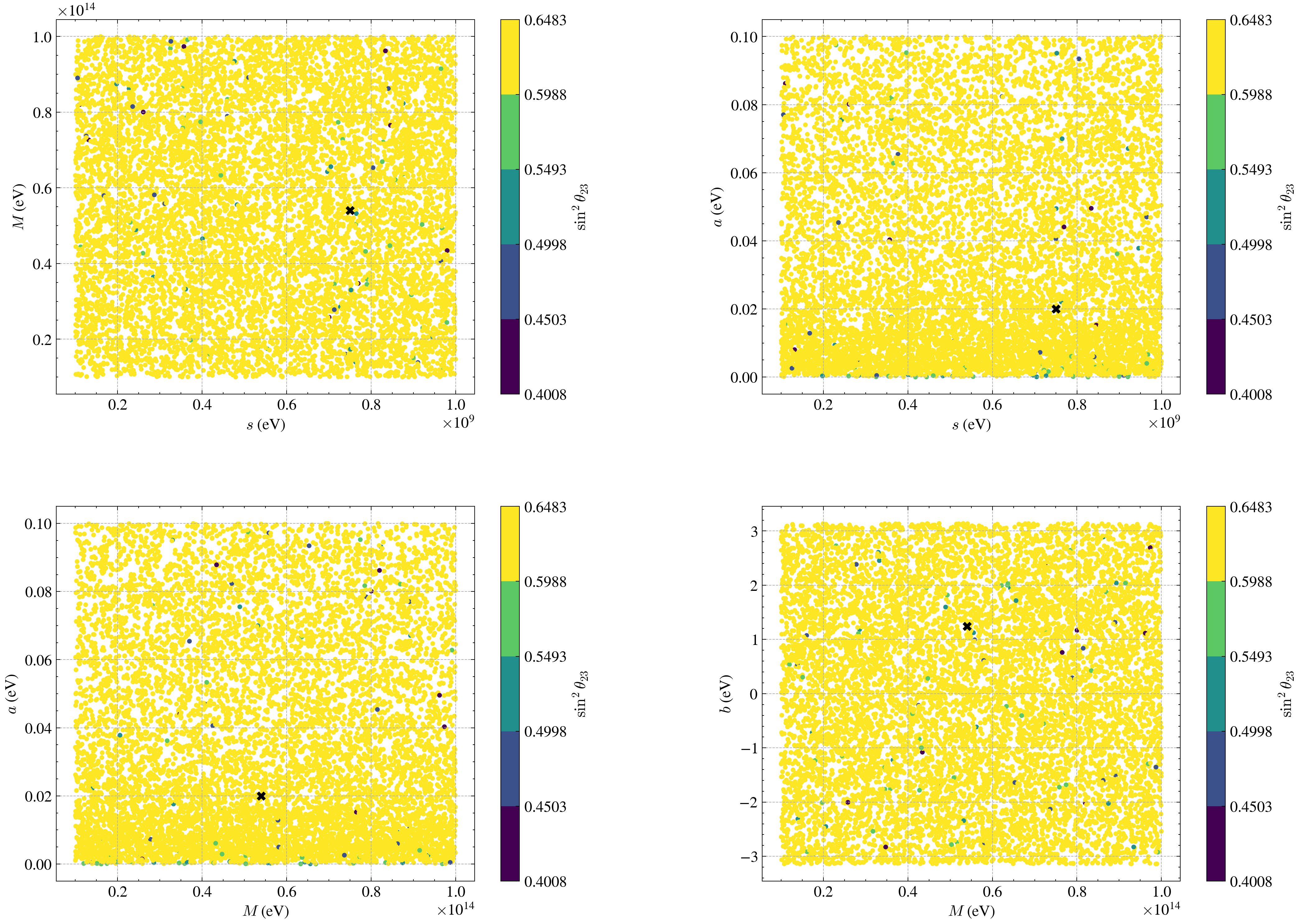}
    \caption{Correlation between the parameters $a$, $b$, $s$ and $M$. The best fit value is
indicated by the $\mathbf{x}$ marker corresponding to $\chi^2_{min}$.}
    \label{fig:1}
\end{figure}

A constraint is applied   on
the sum of absolute neutrino masses	from the cosmological bound $\sum_{i} m_i < 0.12 eV$. In our study, the four complex parameters of the model
are treated as free parameters and are allowed to run over the
following ranges:
$\lvert a \rvert \in [0, 1] eV $, $ \phi_a \in [-\pi , \pi]$ ;
$\lvert b \rvert \in [0, 10^{-1}] eV$ ; $\phi_b \in [-\pi , \pi]$. The two real parameters are allowed to run over the ranges: $ M \in [10^{13}, 10^{14}] eV$ and $ s \in [10^{8}, 10^{9}] eV$.  \\
The term $\phi_a$ and $\phi_b$ are the phases.

\begin{figure}[t]
     \centering
     \begin{subfigure}[b]{0.42\textwidth}
         \centering         \includegraphics[width=\textwidth]{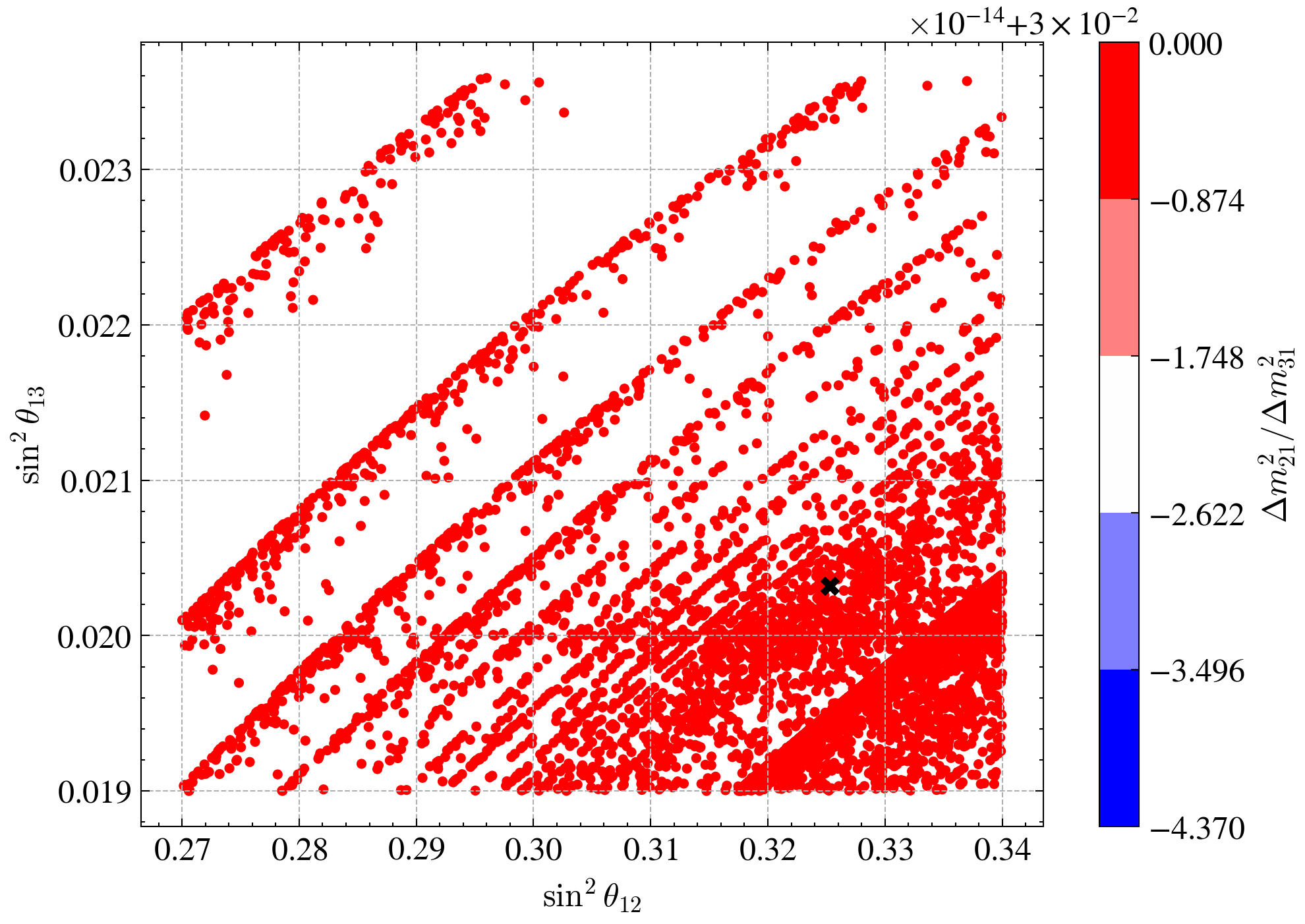}
     \end{subfigure}
     \hfill
     \begin{subfigure}[b]{0.42\textwidth}
         \centering
         \includegraphics[width=\textwidth]{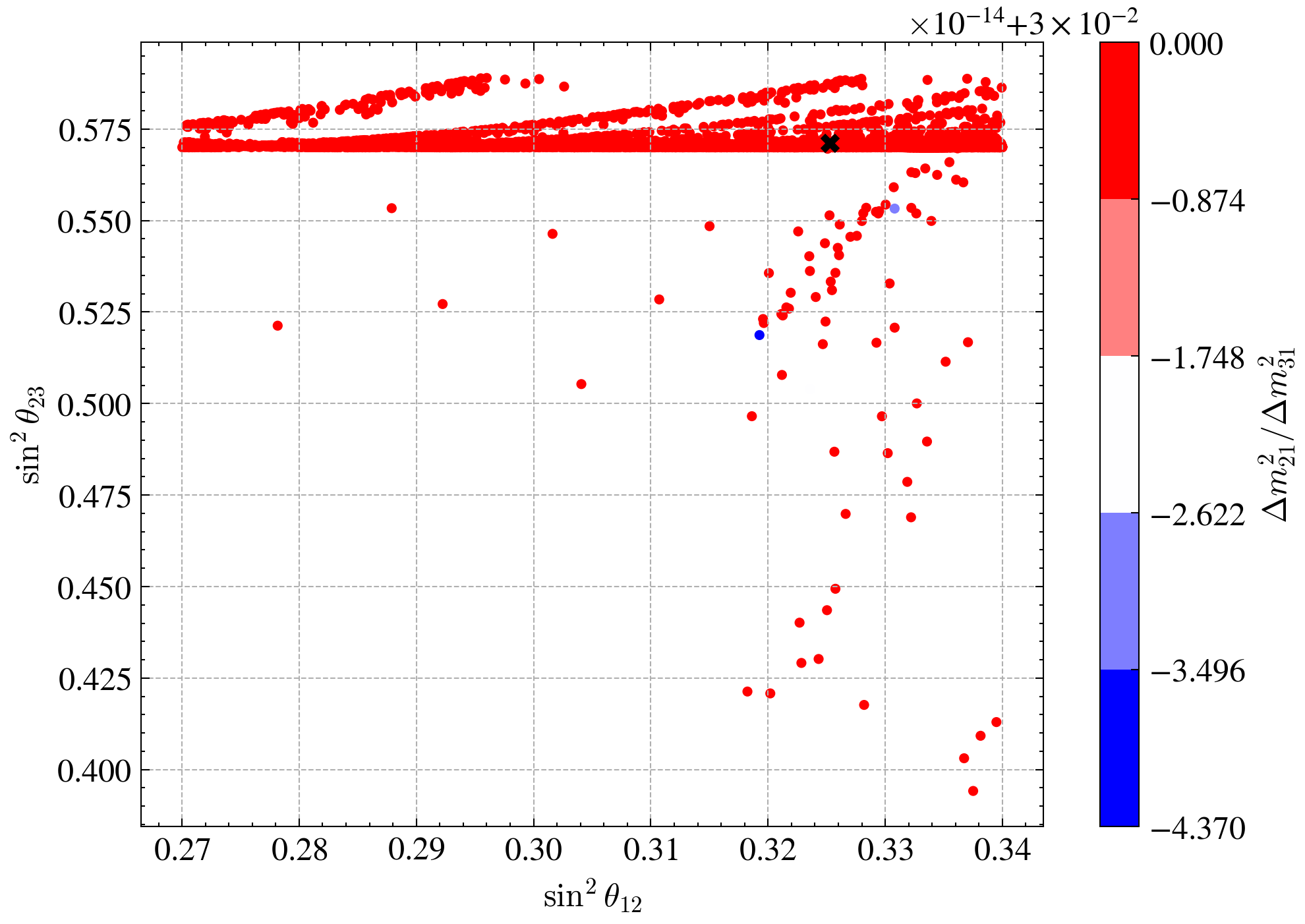}
     \end{subfigure}
     \vspace{1em}
     \begin{subfigure}[b]{0.42\textwidth}
         \centering
    \includegraphics[width=\textwidth]{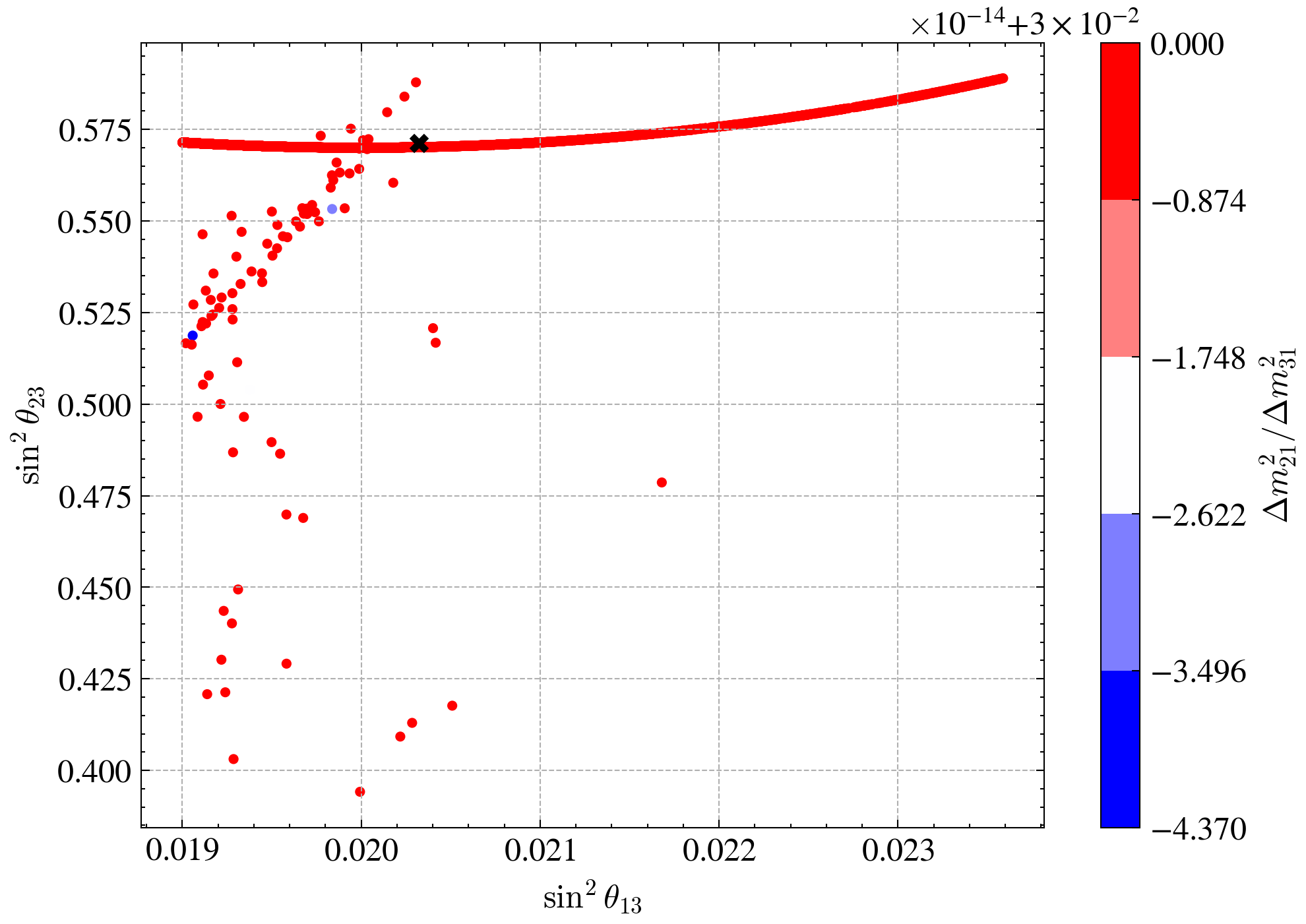}
     \end{subfigure}
    \caption{Correlation among the oscillation parameters predicted by the model at $3\sigma$ in normal hierarchy. The best fit value is
indicated by the $\mathbf{x}$ marker.}
    \label{fig:2}
\end{figure}

 The allowed parameter space of the model is illustrated in Fig.~\ref{fig:1}, subject to all relevant experimental constraints. The best-fit values obtained for the model parameters $a$, $b$, $M$, and $s$ are 0.02, 1.24, $0.54 \times 10^{14}$, and $0.75 \times 10^{9}$, respectively. Figure~\ref{fig:2} presents the predicted values of the neutrino mixing parameters $\sin^2\theta_{12}$, $\sin^2\theta_{13}$, and $\sin^2\theta_{23}$ that provide the best agreement with current experimental observations. The corresponding best-fit values within the experimentally allowed $3\sigma$ ranges are found to be 0.3253, 0.02072, and 0.57121, respectively.
It is observed that the atmospheric mixing angle, $\sin^2\theta_{23}$, is slightly greater than 0.5, indicating that $\theta_{23}$ lies in the upper octant. The model successfully reproduces the observed neutrino oscillation data while remaining consistent with present experimental bounds. Furthermore, the best-fit values of the mass-squared difference ratio, $\Delta m^2_{21}/\Delta m^2_{31}$, correspond to the minimum value of the $\chi^2$ function. Overall, the numerical analysis shows that the model provides a viable framework for explaining the observed neutrino mixing pattern and mass \cite{bora2023neutrino}.

\begin{figure}[ht]
    \centering
    \includegraphics[width=1\linewidth]{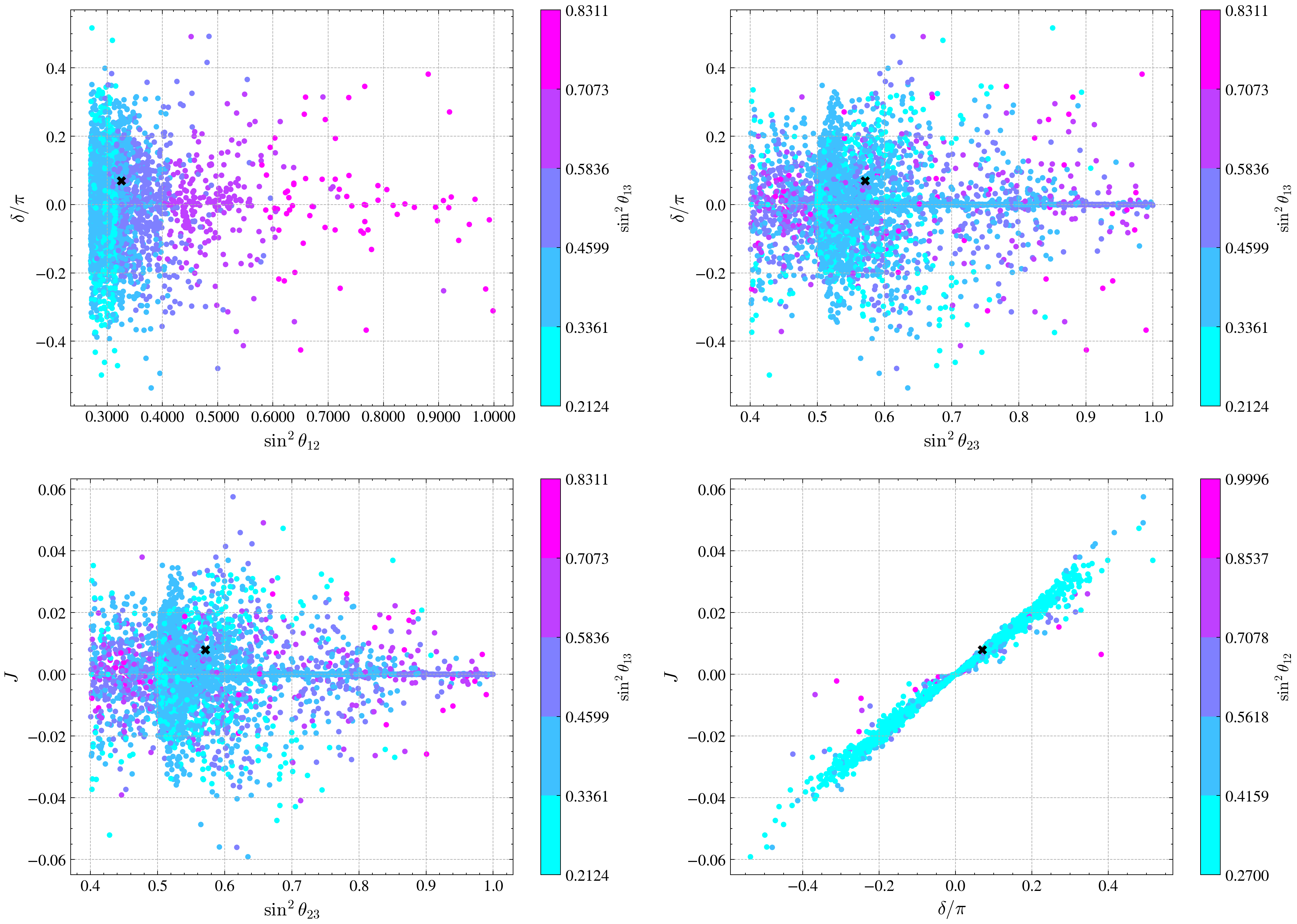}
    \caption{Correlation between Dirac CP ($\delta_{CP}$ ) with  solar and atmospheric mixing angle  respectively (top row).  Correlation between Jarlskog invariant ($J$) with atmospheric mixing angle and $\delta_{CP}$ respectively (bottom row). The best fit value is
indicated by the $\mathbf{x}$ marker.}
    \label{fig:3}
\end{figure}

\textbf{Prediction for CP-violation and Jarlskog invariant parameter}: Figure~\ref{fig:3} presents further predictions of the revised model related to CP violation and the Jarlskog invariant parameter ($J$), which plays an important role in the study of leptonic CP violation and neutrino phenomenology. The Jarlskog invariant is a rephasing-invariant quantity, meaning that its value remains unchanged under phase transformations of the lepton fields. Therefore, it provides a reliable measure of genuine CP violation in the lepton sector. The analysis of these observables is particularly significant for understanding the origin of the matter--antimatter asymmetry in the Universe through leptogenesis mechanisms. 
\begin{equation}
    J = Im(U_{11}U_{22}U^*_{12}U^*_{21}) = s_{12}c^2_{13}s_{12}c_{12}s_{23}c_{23}sin\delta
\end{equation}
The numerical analysis of the model predicts the best-fit value of the Dirac CP-violating phase, $\delta_{CP}$, to be approximately $0.085\pi$, while the corresponding value of the Jarlskog invariant is found to be around 0.0083.

The correlation between the neutrino oscillation parameters makes it obvious that the neutrino mixing differs from the TBM mixing in NH. The upper octant is preferred in the NH scenario, according to the prediction of mixing angle $\theta_{23}$. It is feasible to depart from TBM mixing by modifying the $\Delta(54)$ model.

\newpage

\section{Dark Matter Sector}
\label{DM}
A non-resonantly created sterile dark matter (DM) may be easily accommodated within an extended SM of particle
physics. Since the mass of the sterile neutrino is limited to a few keV, it is a strong candidate for warm dark matter.
The only way the sterile neutrinos interact with the particles of the standard model is via mixing with the active
neutrinos. Thus, the DM abundance can ultimately build up from the active neutrinos that are a part of the
original plasma because they interact weakly. This mechanism is known as the Dodelson and Widrow (DW)
mechanism \cite{kang2016sterile,palazzo2007sterile,bora2024relic}.

When lepton asymmetry is absent, sterile neutrino non-resonant production (NRP) occurs. On the other hand, resonant sterile neutrinos with tiny mixing angles can be produced for a convincing amount of lepton asymmetry in
the original plasma, leading to noticeably cooler momenta. Another name for this process is the resonant production
mechanism, or Shi and Fuller (SF) \cite{shi1999new} . Non-resonant production accounts for the smallest amount of dark matter
contribution that can be created as a result of the dark matter mass and the mixing angle. As a non-resonant DM
candidate, we have taken into account a sterile neutrino mass matrix, in our model. Therefore $m_{DM}$ will be used to represent the fermion mass\cite{bora2024relic}.

The lightest sterile fermion, dominantly composed of the gauge-singlet fields $S_L$ and $S_R$
, forms a pseudo-Dirac state with suppressed active–sterile mixing and serves as a viable dark matter candidate in the triple inverse seesaw \cite{bora2024relic}.

\begin{figure}[t]
      \centering
\includegraphics[width=0.5\linewidth]{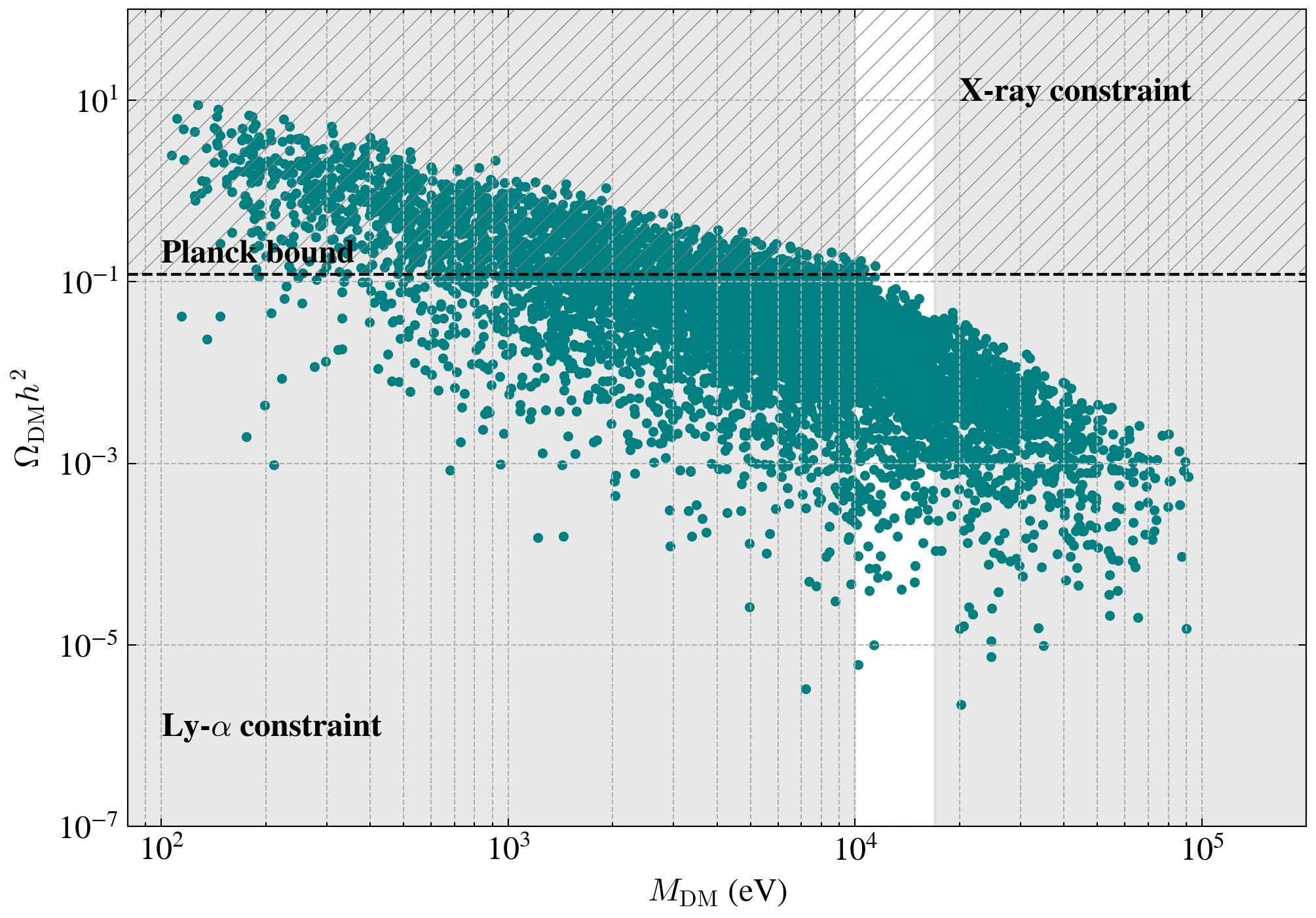}
      \caption{Correlation between Relic abundance ($\Omega_{DM}h^2$) as a function of dark matter
mass ($M_{DM}$) and including constraints from Lyman-$\alpha$ and X-ray for NH.}
      \label{fig:6}
  \end{figure}

We solve the  
of the model with some fixed values of variables such as the $M$
in the range $10^{13}- 10^{14}$ GeV. Additionally, the variable $s$ is considered with a mass between $10^{8}$ and $10^{9}$ eV. By calculating these values, we are able to obtain the mixing angle $Sin^2(2\theta_{DM})$ that satisfies the cosmological constraints as well as the required mass $m_{DM}$ in the keV range. In our study, the non-vanishing components  $V_{14}$ and $V_{34}$ of the mixing matrix V, contribute to the mixing angles. The relic abundance simplified equation for non-resonantly generated dark matter takes the form as \cite{abazajian2001sterile, bora2024relic}
\begin{equation}
   \Omega_{DM} h^2 \simeq 0.3 \times 10^{10} Sin^2(2\theta_{DM})\bigg(\frac{M_{DM} \times 10^{-2}}{\text{keV}}\bigg)^2
\end{equation}
where, $\Omega_{DM}$ is directly proportional to $m_{DM}$ which is the DM mass as mentioned earlier
and $Sin^2(2\theta_{DM})$ viz the active-DM mixing angle with $Sin^2(2\theta_{DM}) = 4(V^2_{14} + V^2_{34})$. Here, $V^2_{14}$ and $V^2_{34}$ are the $14$ and $34$ element of the $V$ matrix in Eq.\ref{eq8}. In Fig.\ref{fig:6} we have
shown the dark matter mass range which obeys the Planck limit for
relic abundance of dark matter. We have larger number of points for in
the allowed parameter space $M_{DM}$ = $10-16$ keV satisfying the Planck
bound.\\
The Sterile mass is obtained as \cite{bora2024relic}:
\begin{equation}
    m_s = \frac{s^2}{M}
\end{equation}
which is considered as the dark matter mass and is represented as $M_{DM}$ in our analysis.

\begin{figure}[t]
    \centering
    \includegraphics[width=1\textwidth]{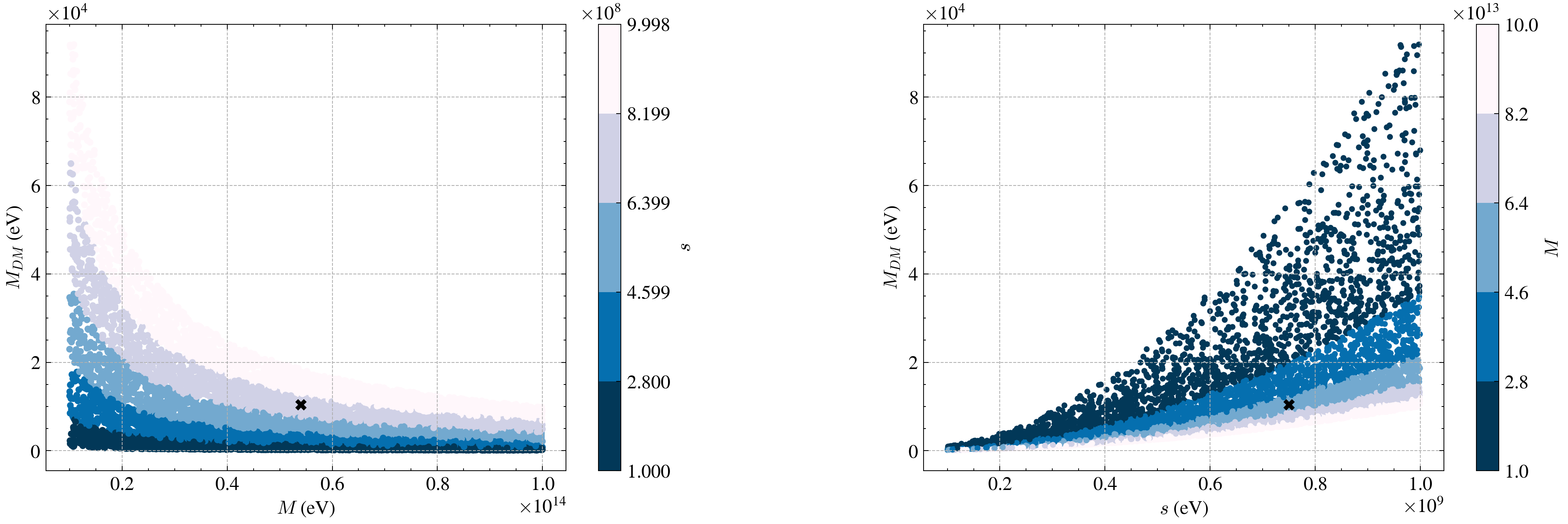}
    \caption{Allowed regions of the parameter $M_{DM}$. The best fit value is
indicated by the $\mathbf{x}$ marker.}
    \label{fig:4}
\end{figure}

 \begin{figure}[t]
      \centering
\includegraphics[width=0.5\linewidth]{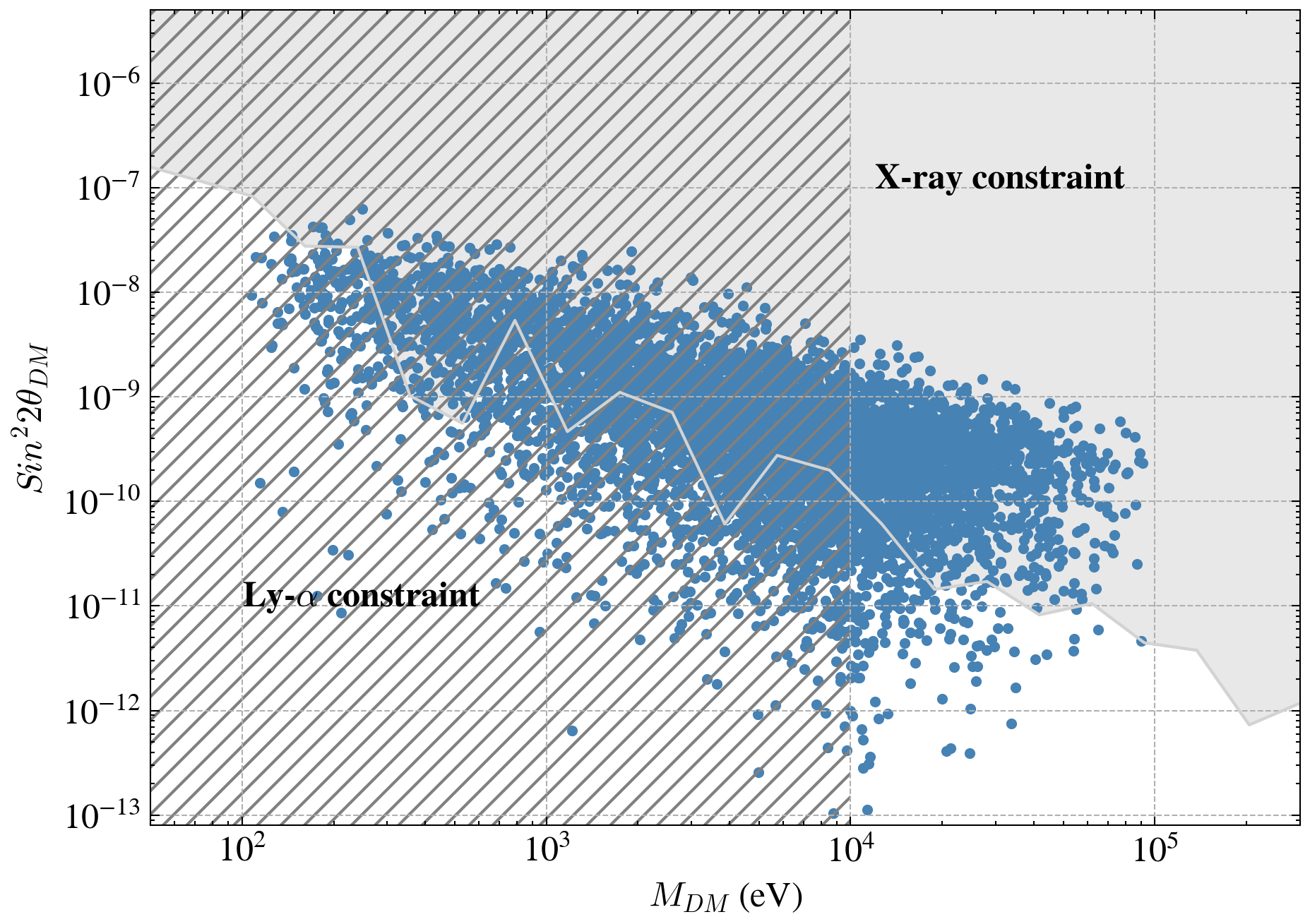}
      \caption{Correlation between Active neutrino-DM mixing angle $\textit{Sin}^2 2\theta_{DM}$ as a function
of dark matter mass including constraints from Lyman-$\alpha$ and X-ray for NH.}
      \label{fig:5}
  \end{figure}

\newpage  
\section{Resonant Leptogenesis} 
\label{Res}
Fukugita and Yanagida originally proposed the leptogenesis mechanism, which is one of the most commonly accepted explanations for the Baryon Asymmetry of the Universe (BAU). The mass of the lightest right-handed neutrino, $M_1 = 10^9$ GeV, has a lower bound in the simplest case of thermal leptogenesis with a hierarchical mass spectrum of right-handed neutrinos \cite{davidson2002lower}. Although one can lower this limit if their masses are nearly degenerate. This scenario is
popularly known as resonant leptogenesis\cite{pilaftsis2004resonant, pilaftsis1997cp}. In this scenario, the resonant enhancement amplifies the one-loop self-energy contribution, leading to the flavor-dependent asymmetry resulting from the decay of a right-handed neutrino into a lepton and Higgs \cite{xing2020bridging, bora2026neutrino}.
\begin{equation}
\epsilon_{i\alpha} = \frac{{\Gamma(N_i \to l_\alpha + H) - \Gamma(N_i \to \bar{l}_\alpha + \bar{H})}}{{\sum_{\alpha} \left(\Gamma(N_i \to l_{\alpha} + H) + \Gamma(N_i \to \bar{l}_{\alpha} + \bar{H})\right)}} 
\end{equation}

\begin{equation}
\label{eq:3}
= \sum_{i \neq j}\frac{{  \text{Im}\biggl\{(Y^*_{\nu})_{ \alpha i} (Y_{\nu})_{\alpha j} \bigr[(Y_{\nu}^{\dagger} Y_{\nu})_{ij} + \xi_{ij} (Y_{\nu}^{\dagger} Y_{\nu})_{ji}}\bigr]\biggl\}} {(Y_{\nu}^{\dagger} Y_{\nu})_{ii} (Y_{\nu}^{\dagger} Y_{\nu})_{jj} } \times \frac{\xi_{ij} \zeta_j (\xi_{ij}^2 - 1)}{ (\xi_{ij} \zeta_j)^2 + (\xi_{ij}^2 - 1)^2}  
\end{equation}\\
where $\xi_{ij} = M_i/M_j$ and we took  $M_1 = 10 $ TeV  and $d = (M_3 - M_1)/M_1 = 10^{-8}$. The plots of correlation between the flavor-dependent CP-violating asymmetries are shown in Fig.\ref{fig:6}.\\
In the type-I seesaw, the effective neutrino mass matrix of Eq.\ref{eq8} can be written as
\begin{align} 
m_\nu =& M_{\nu N}M^{-1}_{mid}M^\prime_{\nu N} 
  \quad \text{with} \quad M_{mid} = \frac{M^{\prime 2}_{NS} M^{\prime}_{\mu}}{M^2_N M^2_{S}}
\end{align} \\
The Majorana mass matrix $M_{mid}$ for the right-handed neutrinos possesses three degenerate mass eigenvalue, $M_1 = M_2 = M_3 = M$.
\begin{equation}
     M_{mid}=
    \begin{pmatrix}
     M  &  0  &  0\\
     0 &  M & 0 \\
     0 &   0  & M
    \end{pmatrix} 
    \label{eq28}
\end{equation}
\begin{figure}[t]
    \centering
    \includegraphics[width=1\linewidth]{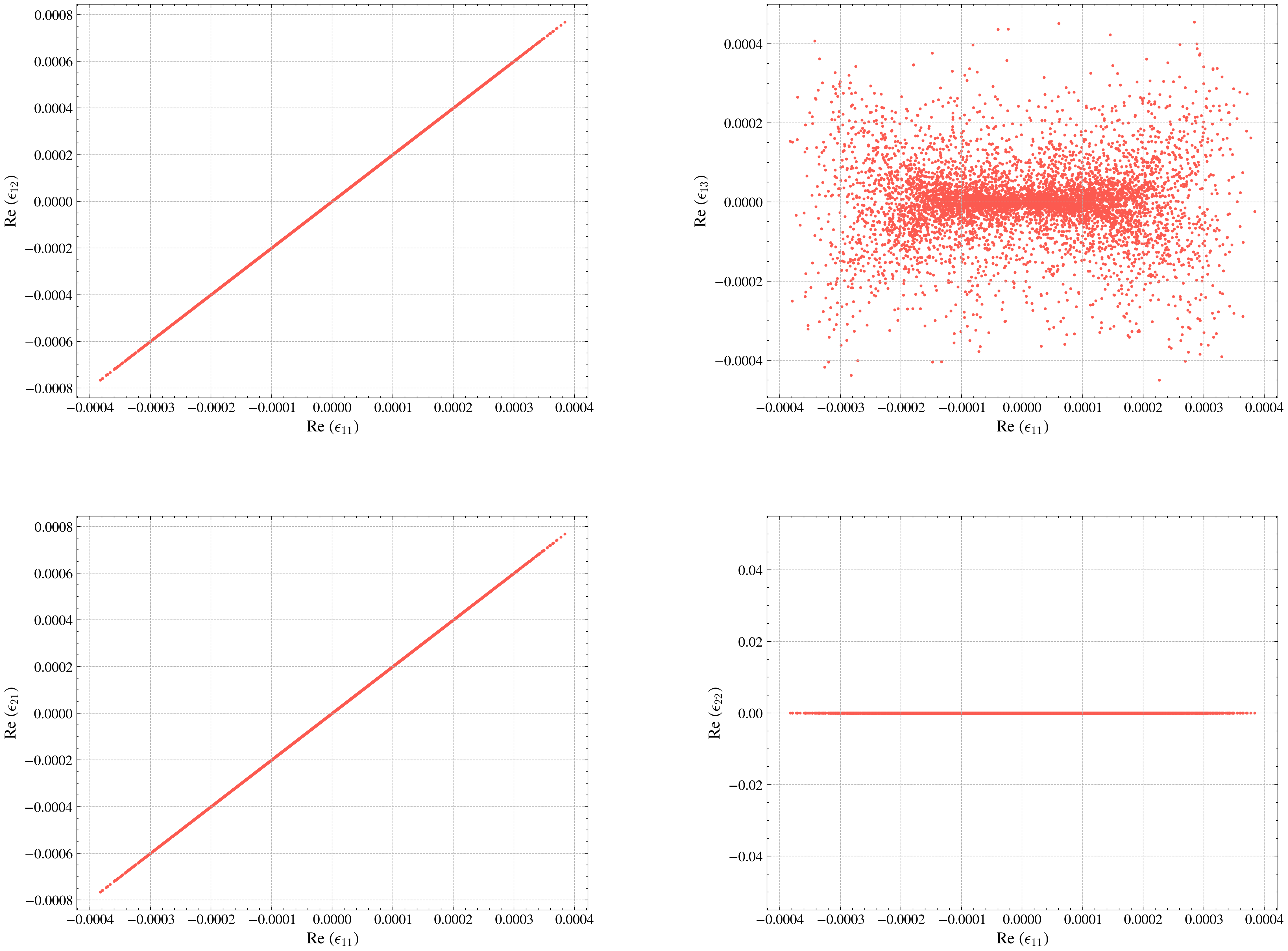}
    \caption{Correlations between the flavor-dependent CP-violating asymmetries $\epsilon_{11}$ with $\epsilon_{12}$, $\epsilon_{21}$ and $\epsilon_{22}$.
respectively.}
    \label{fig:7}
\end{figure}
where, $M= \frac{v^{\prime 2}_\zeta y^{\prime 2}_{NS} y^\prime_\mu v^{\prime \prime}_\phi}{v^2_\phi y^2_N y^2_S v^{\prime 2}_\phi} $. However, a tiny mass separation between any two right-handed neutrinos is necessary for successful leptogenesis, and this is included to our model by including a higher dimension term, $y_{\mu2}^{\prime} S_L\Bar{S_L}\phi^{\prime\prime}$ in our Lagrangian \cite{thapa2021resonant}. Such term leads to a minor correction in the Majorana mass matrix of Eq.\ref{eq8}, and the resultant structure of the mass matrix may be \cite{bora2026neutrino}as \footnote{For details on the origin of the term $e$ from the $\Delta(54)$ symmetry, see Appendix.}
\begin{equation}
     M_{mid}=
    \begin{pmatrix}
     M  &  e  &  e\\
     e  &  M  & e  \\
     e &   e   & M
    \end{pmatrix} 
    \label{eq29}
\end{equation}
\begin{figure}[t]
    \centering
    \includegraphics[width=1\linewidth]{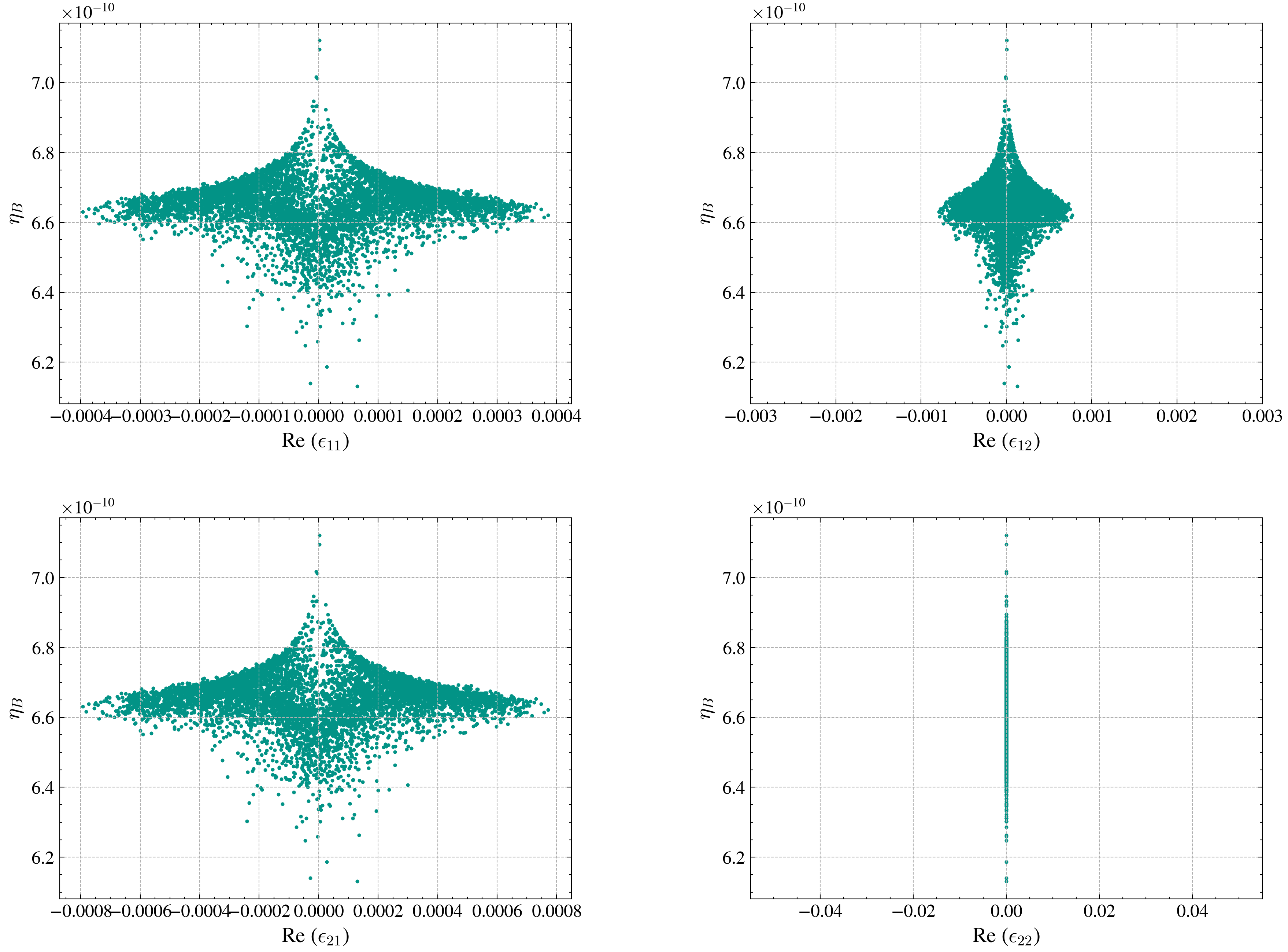}
    \caption{Correlations between the Baryon Asymmetry ($\eta_B$) with flavor-dependent CP-violating asymmetries
$\epsilon_{11}$ , $\epsilon_{12}$, $\epsilon_{21}$ and $\epsilon_{22}$ respectively.}
    \label{fig:8}
\end{figure}
where $  e = y_{\mu2} v^{\prime\prime}_\phi $ is a parameter that quantifies the tiny
difference between masses required for leptogenesis. The
mass matrix in Eq.\ref{eq29} is diagonalized using a ($3\times3$) matrix\cite{bora2026neutrino}
of the form
\begin{equation}
  D = \begin{pmatrix}
    -1  &  -1  &    1\\
     1  &  0 & 1  \\
    0 &   1   & 1 
    \end{pmatrix} 
\end{equation}
with real eigenvalues $M_1 = M - e$ and $M_2 = M - e$ and $M_3 = M - 2e$         . In the
basis where the charged-lepton and Majorana mass matrix
are diagonal, the dirac mass matrix Eq.(\ref{eq6}) takes the form
\begin{equation}
M^\prime_{\nu N} = M_{\nu N}.D = 
    \begin{pmatrix}
    \frac{v(b-a)}{\Lambda}  &  \frac{v(b-a)}{\Lambda} &    \frac{v(2b+a)}{\Lambda}\\
     \frac{v(-b+a)}{\Lambda}  &  0 & \frac{v(2b+a)}{\Lambda} \\
    0 &   \frac{v(-b+a)}{\Lambda}   & \frac{v(2b+a)}{\Lambda} 
    \end{pmatrix}  
\end{equation}
From this point onward, we will take $Y_{\nu N} = M^\prime_{\nu N} / v $, which is
relevant for calculating CP asymmetry that arises during the
decay of right-handed neutrinos in out-of-equilibrium way.\\
  The CP-violating asymmetries $\varepsilon_{i\alpha}$ in the flavored resonant leptogenesis scenario under study are related to the baryon-to-photon ratio $\eta_B$ as follows \cite{xing2020bridging}:

  \begin{equation}
      \eta_{B} \simeq -9.6 \times 10^{-3} \sum_{\alpha}(\varepsilon_{1\alpha}K_{1\alpha} + \varepsilon_{2\alpha}K_{2\alpha})
  \end{equation}
where $K_{1\alpha}$ and $K_{2\alpha}$ are the conversion efficiency factors. The region in which the lepton flavor takes effect determines the sum over the flavor index $\alpha$. To evaluate the sizes of $K_{i\alpha}$, let us first of all figure out the effective light neutrino masses as
\begin{equation}
    m_{i\alpha} \simeq \frac{v^2 \big\lvert (Y_v)_{\alpha i} \big\rvert^2}{M_i}
    \end{equation}
     The decay parameters $K_{i\alpha} \equiv \Tilde{m}_{i\alpha}/m_*$ can be calculated, where $m_* = 8\pi v^2 H(M_1)/M^2_1 \simeq 1.08 \times 10^{-3}eV$ gives the equilibrium neutrino mass and $H(M_1)$ is called the Hubble expansion parameter of the Universe.\\
    Now we define a dimensionless parameter $d \equiv (M_2 - M_1) / M_1 = \xi_{21} - 1$ to calculate the level of degeneracy for two of the three heavy Majorana neutrinos. Allowing for $d << 1$, we have $\kappa_{1\alpha} \simeq \kappa_{2\alpha} \equiv \kappa(K_\alpha)$ with $K_{\alpha} \equiv K_{1\alpha} + K_{2\alpha}$. Given the initial thermal abundance of heavy Majorana neutrinos, the efficiency factor $\kappa(K_\alpha)$ can be expressed as
\begin{equation}
   \kappa(K_\alpha) \simeq \frac{2}{K_\alpha z_B(K_\alpha)} \biggr[ 1 - exp \biggl( \frac{-1}{2} K_\alpha z_B(K_\alpha) \biggl) \biggr]  
\end{equation}
where $z_B(K_\alpha) \simeq 2 + 4 K^{0.13}_\alpha exp(-2.5 / K_\alpha)$.  We have the correlation of Baryon Asymmetry of the Universe with the flavor dependent CP-violating asymmetries in Fig.\ref{fig:8}.

\section{Conclusion}
\label{conc}
In this work, we have presented a flavor model based on the non-Abelian discrete symmetry group $\Delta(54)$, which is distinguished by the presence of multiple non-trivial field contractions. To construct a viable neutrino mass matrix, the $\Delta(54)$ symmetry is supplemented by an auxiliary $Z_2\otimes Z_3\otimes Z_4$ symmetry. Neutrino masses are generated through a triple inverse seesaw mechanism, providing a flavor-symmetric framework that successfully reproduces the observed neutrino masses and mixing parameters.

The model predicts a non-zero reactor mixing angle $\theta_{13}$  within the allowed parameter space. Furthermore, the atmospheric mixing angle $\theta_{23}$ shows a preference for the upper octant.  We have also examined the CP-violating phase ($\delta_{CP}$) and the Jarlskog invariant parameter ($J$), obtaining values of 0.085$\pi$ and 0.0083, respectively. A numerical $\chi^2$ analysis demonstrates that the predicted neutrino oscillation parameters are in good agreement with the current global best-fit values for the Normal Hierarchy (NH), while the Inverted Hierarchy (IH) scenario is disfavored by experimental data.

We further identified a sterile neutrino arising from the model as a viable dark matter (DM) candidate and computed both its mass and the active–sterile mixing angle using the predicted model parameters. The non-resonant production mechanism of sterile neutrino dark matter and the associated Lyman-$\alpha$ constraints were carefully taken into account. We analyzed the correlation between the dark matter mass and the active–DM mixing angle and found that, for the NH case, the allowed parameter space favors the Lyman-$\alpha$ bound. Specifically, a narrow mass window in the range $m_{\rm DM} \simeq 10$–$16$ keV satisfies both the Lyman-$\alpha$ constraint, with a lower bound of $10$ keV, and the X-ray constraint, which imposes a lower limit of approximately $17$ keV \cite{bora2024relic}. The viable parameter space is illustrated in Fig.~\ref{fig:5}. Additionally, in Fig.~\ref{fig:6}, we show the region of dark matter mass that satisfies the Planck limit on the relic abundance, with a substantial number of points in the range $m_{\rm DM} = 10$–$16$ keV consistent with the observed dark matter density.

Finally, we investigated the generation of the baryon asymmetry of the Universe through flavored resonant leptogenesis. In our model, the right-handed neutrinos are degenerate at the dimension-five level, while higher-dimensional operators induce a small mass splitting. Taking the splitting parameter to be $d \simeq 10^{-8}$, we obtained a resonantly enhanced CP asymmetry $\epsilon_{i\alpha}$ from the out-of-equilibrium decay of right-handed Majorana neutrinos. Using these CP asymmetries, we computed the resulting baryon-to-photon ratio $\eta_B$ and found that the observed value of the baryon asymmetry can be successfully explained for a right-handed neutrino mass scale of $M_1 = 10$ TeV and mass splitting $d \simeq 10^{-8}$.

The extension of the present framework to higher-order Majorana inverse seesaw realizations such as quadruple, quintuple, or even higher-order schemes is straightforward. The key requirement is that the $n$th-order inverse seesaw contribution remains dominant in generating neutrino masses, which can be naturally achieved by assigning appropriate symmetry charges to the fields responsible for the $\mu$-terms. Such extensions are expected to give rise to distinctive phenomenological signatures, potentially observable in collider experiments as well as in low-energy searches for lepton flavor and lepton number violation. The present study also provides a solid foundation for exploring leptogenesis and asymmetric dark matter scenarios within this framework. The model can be tested in current and upcoming neutrino oscillation and dark matter experiments such as DUNE, JUNO, Super-Kamiokande, and related searches, making it a phenomenologically rich and experimentally verifiable framework for explaining neutrino masses, dark matter, and the baryon asymmetry of the Universe.


 \section*{Appendix}
 \subsection*{A.1 \quad Character table of $\Delta(54)$}

 The $\Delta(54)$ is a discrete subgroup of SU(3) and it has order 54. The generartors of $\Delta(54)$ are given by the set\\
 \begin{equation*}
      a=
    \begin{pmatrix}
     0  &  1  &  0\\
     0  & 0  & 1  \\
     1 &   0   & 0
    \end{pmatrix} , \quad
    b =   \begin{pmatrix}
     0  &  1  &  0\\
     0  & 0  & 1  \\
     1 &   0   & 0
    \end{pmatrix}, \quad
    c=   \begin{pmatrix}
     0  &  1  &  0\\
     0  & 0  & 1  \\
     1 &   0   & 0
    \end{pmatrix}, \quad
    c^{\prime}=   \begin{pmatrix}
     0  &  1  &  0\\
     0  & 0  & 1  \\
     1 &   0   & 0
    \end{pmatrix}
 \end{equation*}

The $\Delta(54)$ symmetry has four three-dimensional irreducible representations
$3_{1(1)}, 3_{1(2)}, 3_{2(1)}, 3_{2(2)}$, four two-dimensional
ones $2_1, 2_2, 2_3, 2_4$, 
and two one-dimensional ones $1_1, 1_2$. 
Generators of three
dimensional
representations are mainly divided into two types. For $3_{1(1)}, 3_{1(2)}$, generators are $a, b, c$ and for $3_{2(1)}, 3_{2(2)}$
are $a, b, c^{\prime}$. Their character table are presented in Table.\ref{tab:3}
\begin{table}[ht]
    \centering
 \begin{tabular}{c c c c c c c c c c c}
    \hline
      &  $1a$ & $6a$ & $6b$ & $3a$  & $3b$ & $3c$  & $2a$& $3d$ &$ 3e$ & $3f$ \\
     \hline
    $ 1_{1}$  &  1 & 1 & 1 & 1  & 1 & 1  & 1& 1 & 1 & 1 \\
     \hline
     $1_2$  &  1 & -1 & -1 & 1  & 1 & 1  & -1& 1 & 1 & 1 \\
     \hline
     $2_1$  &  2 & 0 & 0 & 2  & -1 & -1  & 0& -1 & 2 & 2 \\
     \hline
 $2_2$  &  2 & 0 & 0 & -1  & -1 & -1  & 0& 2 & 2 & 2 \\
     \hline
 $2_3$  &  2 & 0 & 0 & -1  & -1 & 2  & 0& -1 & 2 & 2 \\
 \hline
  $2_4 $ &  2 & 0 & 0 & -1  & 2 & -1  & 0& -1 & 2 & 2 \\
     \hline
   $ 3_{2(1)}$  &  3 & $-\bar{\omega}$ & $-\omega$ & 0  & 0 & 0 & -1& 0 & $3\bar{\omega}$ & $3\omega$ \\
     \hline
      $3_{2(2)} $ &  3 & $-\omega$  &  $-\bar{\omega}$ & 0 & 0 & 0 & -1 & 0 & $3\omega$ & $3\bar{\omega}$ \\
     \hline
      $3_{1(2)}$  &  3 &  $\omega$ &  $\bar{\omega}$ & 0  & 0 & 0 & 1 & 0 & $3\omega$ & $3\bar{\omega}$ \\
     \hline
      $3_{1(1)}$ &  3 &  $\bar{\omega}$ &  $\omega$ & 0  & 0 & 0 & 1& 0& $3\bar{\omega}$ & $3\omega$ \\
     \hline
    \end{tabular}
    \caption{Character table of $\Delta(54)$}
    \label{tab:3}
    \end{table}

\subsection*{A.2 \quad Multiplication of $\Delta(54)$}
The irreducible representations of $\Delta(54)$ follow the following Kronecker products,
\begin{align*}
    &1_{1}\otimes S_{i} = S_{i}, \qquad  1_{2}\otimes 1_{2} = 1_{1}, \qquad 1_{2}\otimes 3_{1(1)} = 3_{2(1)} \\
&1_{2}\otimes 3_{1(2)} = 3_{2(2)},\qquad  
1_{2}\otimes 3_{2(1)} = 3_{1(1)} , \qquad 
 1_{2}\otimes 3_{2(2)} = 3_{1(2)} 
\end{align*}
Now, we write the Clebsch-Gordon coefficients in particular basis. The multiplication rules of two dimensional representation is given as: \\
 \begin{align*}
    \begin{pmatrix}
    a_1\\ a_2 \end{pmatrix}_{2_{s}} \otimes   \begin{pmatrix}
    b_1\\ b_2 \end{pmatrix}_{2_{s}} = \begin{pmatrix}
    a_1 b_2 +  a_2 b_1 \end{pmatrix}_{1_{1}} \oplus \begin{pmatrix}
    a_1 b_2 -  a_2 b_1\end{pmatrix}_{1_{2}} \oplus \begin{pmatrix}
    a_2 b_2 \\ a_1 b_1\end{pmatrix}_{2_{s}}
\end{align*}
\begin{align*}
    \begin{pmatrix}
    a_1\\ a_2\end{pmatrix}_{2_{1}} \otimes   \begin{pmatrix}
    b_1\\b_2\end{pmatrix}_{2_{2}} = \begin{pmatrix}
    a_{2}b_{2}\\a_{1}b_{1}\end{pmatrix}_{2_{3}}  \oplus \begin{pmatrix}
     a_{2}b_{1}\\a_{1}b_{2} \end{pmatrix}_{2_{4}} 
\end{align*}
\begin{align*}
    \begin{pmatrix}
    a_1\\ a_2\end{pmatrix}_{2_{1}} \otimes   \begin{pmatrix}
    b_1\\b_2\end{pmatrix}_{2_{3}} = \begin{pmatrix}
    a_{2}b_{2}\\a_{1}b_{1}\end{pmatrix}_{2_{2}}  \oplus \begin{pmatrix}
     a_{2}b_{1}\\a_{1}b_{2} \end{pmatrix}_{2_{4}} 
\end{align*}
\begin{align*}
    \begin{pmatrix}
    a_1\\ a_2\end{pmatrix}_{2_{1}} \otimes   \begin{pmatrix}
    b_1\\b_2\end{pmatrix}_{2_{4}} = \begin{pmatrix}
    a_{1}b_{2}\\a_{2}b_{1}\end{pmatrix}_{2_{2}}  \oplus \begin{pmatrix}
     a_{1}b_{1}\\a_{2}b_{2} \end{pmatrix}_{2_{3}} 
\end{align*}
\begin{align*}
    \begin{pmatrix}
    a_1\\ a_2\end{pmatrix}_{2_{2}} \otimes   \begin{pmatrix}
    b_1\\b_2\end{pmatrix}_{2_{3}} = \begin{pmatrix}
    a_{2}b_{2}\\a_{1}b_{1}\end{pmatrix}_{2_{1}}  \oplus \begin{pmatrix}
     a_{1}b_{2}\\a_{2}b_{1} \end{pmatrix}_{2_{4}} 
\end{align*}
\begin{align*}
    \begin{pmatrix}
    a_1\\ a_2\end{pmatrix}_{2_{2}} \otimes   \begin{pmatrix}
    b_1\\b_2\end{pmatrix}_{2_{4}} = \begin{pmatrix}
    a_{1}b_{1}\\a_{2}b_{2}\end{pmatrix}_{2_{1}}  \oplus \begin{pmatrix}
     a_{1}b_{2}\\a_{2}b_{1} \end{pmatrix}_{2_{3}} 
\end{align*}
\begin{align*}
    \begin{pmatrix}
    a_1\\ a_2\end{pmatrix}_{2_{3}} \otimes   \begin{pmatrix}
    b_1\\b_2\end{pmatrix}_{2_{4}} = \begin{pmatrix}
    a_{1}b_{2}\\a_{2}b_{1}\end{pmatrix}_{2_{1}}  \oplus \begin{pmatrix}
     a_{1}b_{1}\\a_{2}b_{2} \end{pmatrix}_{2_{2}} 
\end{align*}

The multiplication rules of three dimensional representation is given as:\\
 \begin{align*}
    \begin{pmatrix}
    a_1\\ a_2\\a_3\end{pmatrix}_{3_{1(1)}} \otimes   \begin{pmatrix}
    b_1\\b_2\\b_3\end{pmatrix}_{3_{1(1)}} = \begin{pmatrix}
    a_{1}b_{1}\\a_{2}b_{2}\\a_{3}b_{3}\end{pmatrix}_{3_{1(2)}}  \oplus \begin{pmatrix}
    a_{2}b_{3} + a_{3}b_{2} \\a_{3}b_{1} + a_{1}b_{3} \\a_{1}b_{2} + a_{2}b_{1} \end{pmatrix}_{3_{1(2)}} \oplus  \begin{pmatrix}
    a_{2}b_{3} - a_{3}b_{2} \\a_{3}b_{1} - a_{1}b_{3} \\a_{1}b_{2} - a_{2}b_{1}\end{pmatrix}_{3_{2(2)}}   
\end{align*}
  \begin{align*}
    \begin{pmatrix}
    a_1\\ a_2\\a_3\end{pmatrix}_{3_{1(2)}} \otimes   \begin{pmatrix}
    b_1\\b_2\\b_3\end{pmatrix}_{3_{1(2)}} = \begin{pmatrix}
    a_{1}b_{1}\\a_{2}b_{2}\\a_{3}b_{3}\end{pmatrix}_{3_{1(1)}}  \oplus \begin{pmatrix}
    a_{2}b_{3} + a_{3}b_{2} \\a_{3}b_{1} + a_{1}b_{3} \\a_{1}b_{2} + a_{2}b_{1} \end{pmatrix}_{3_{1(1)}} \oplus  \begin{pmatrix}
    a_{2}b_{3} - a_{3}b_{2} \\a_{3}b_{1} - a_{1}b_{3} \\a_{1}b_{2} - a_{2}b_{1}\end{pmatrix}_{3_{2(1)}}   
\end{align*}
 \begin{align*}
    \begin{pmatrix}
    a_1\\ a_2\\a_3\end{pmatrix}_{3_{2(1)}} \otimes   \begin{pmatrix}
    b_1\\b_2\\b_3\end{pmatrix}_{3_{2(1)}} = \begin{pmatrix}
    a_{1}b_{1}\\a_{2}b_{2}\\a_{3}b_{3}\end{pmatrix}_{3_{1(2)}}  \oplus \begin{pmatrix}
    a_{2}b_{3} + a_{3}b_{2} \\a_{3}b_{1} + a_{1}b_{3} \\a_{1}b_{2} + a_{2}b_{1} \end{pmatrix}_{3_{1(2)}} \oplus  \begin{pmatrix}
    a_{2}b_{3} - a_{3}b_{2} \\a_{3}b_{1} - a_{1}b_{3} \\a_{1}b_{2} - a_{2}b_{1}\end{pmatrix}_{3_{2(2)}}   
\end{align*}
 \begin{align*}
    \begin{pmatrix}
    a_1\\ a_2\\a_3\end{pmatrix}_{3_{2(2)}} \otimes   \begin{pmatrix}
    b_1\\b_2\\b_3\end{pmatrix}_{3_{2(2)}} = \begin{pmatrix}
    a_{1}b_{1}\\a_{2}b_{2}\\a_{3}b_{3}\end{pmatrix}_{3_{1(1)}}  \oplus \begin{pmatrix}
    a_{2}b_{3} + a_{3}b_{2} \\a_{3}b_{1} + a_{1}b_{3} \\a_{1}b_{2} + a_{2}b_{1} \end{pmatrix}_{3_{1(1)}} \oplus  \begin{pmatrix}
    a_{2}b_{3} - a_{3}b_{2} \\a_{3}b_{1} - a_{1}b_{3} \\a_{1}b_{2} - a_{2}b_{1}\end{pmatrix}_{3_{2(1)}} 
\end{align*}
 \begin{align*}
     \begin{pmatrix}
    a_1\\ a_2\\a_3\end{pmatrix}_{3_{1(1)}} \otimes   \begin{pmatrix}
    b_1\\b_2\\b_3\end{pmatrix}_{3_{1(2)}} = &\begin{pmatrix}
    a_{1}b_{1} + a_{2}b_{2} + a_{3}b_{3} \end{pmatrix}_{1_{1}}  \oplus \begin{pmatrix}
    a_{1}b_{1} + \omega^2 a_{2}b_{2} + \omega a_{3}b_{3 }\\ \omega a_{1}b_{1} + \omega^2 a_{2}b_{2} +  a_{3}b_{3 }\end{pmatrix}_{2_{1}} \\&
    \oplus  \begin{pmatrix}
    a_{1}b_{2} + \omega^2 a_{2}b_{3} + \omega a_{3}b_{1 }\\\omega a_{1}b_{3} + \omega^2 a_{2}b_{1} + a_{3}b_{2}\end{pmatrix}_{2_{2}}  \oplus  \begin{pmatrix}
    a_{1}b_{3} + \omega^2 a_{2}b_{1} + \omega a_{3}b_{2}\\ \omega a_{1}b_{2} + \omega^2 a_{2}b_{3} + \omega a_{3}b_{1}\end{pmatrix}_{2_{3}} \\& \oplus \begin{pmatrix}
    a_{1}b_{3} +  a_{2}b_{1} + a_{3}b_{2}\\a_{1}b_{2} +  a_{2}b_{3} + a_{3}b_{1 }\end{pmatrix}_{2_{4}} \\
\end{align*}
\begin{align*}
    \begin{pmatrix}
    a_1\\ a_2\\a_3\end{pmatrix}_{3_{1(1)}} \otimes   \begin{pmatrix}
    b_1\\b_2\\b_3\end{pmatrix}_{3_{2(1)}} = \begin{pmatrix}
    a_{1}b_{1}\\a_{2}b_{2}\\a_{3}b_{3}\end{pmatrix}_{3_{2(2)}}  \oplus \begin{pmatrix}
    a_{3}b_{2} - a_{2}b_{3} \\a_{1}b_{3} - a_{3}b_{1} \\a_{2}b_{1} - a_{1}b_{2} \end{pmatrix}_{3_{1(2)}} \oplus  \begin{pmatrix}
    a_{3}b_{2} + a_{2}b_{3} \\a_{1}b_{3} + a_{3}b_{1} \\a_{2}b_{1} + a_{1}b_{2}\end{pmatrix}_{3_{2(2)}}   
\end{align*}
 \begin{align*}
     \begin{pmatrix}
    a_1\\ a_2\\a_3\end{pmatrix}_{3_{1(1)}} \otimes   \begin{pmatrix}
    b_1\\b_2\\b_3\end{pmatrix}_{3_{2(2)}} = &\begin{pmatrix}
    a_{1}b_{1} + a_{2}b_{2} + a_{3}b_{3} \end{pmatrix}_{1_{2}}  \oplus \begin{pmatrix}
    a_{1}b_{1} + \omega^2 a_{2}b_{2} + \omega a_{3}b_{3 }\\ -\omega a_{1}b_{1} - \omega^2 a_{2}b_{2} -  a_{3}b_{3 }\end{pmatrix}_{2_{1}} \\&
    \oplus  \begin{pmatrix}
    a_{1}b_{2} + \omega^2 a_{2}b_{3} + \omega a_{3}b_{1 }\\-\omega a_{1}b_{3} - \omega^2 a_{2}b_{1} - a_{3}b_{2}\end{pmatrix}_{2_{2}}  \oplus  \begin{pmatrix}
    a_{1}b_{3} + \omega^2 a_{2}b_{1} + \omega a_{3}b_{2}\\ -\omega a_{1}b_{2} - \omega^2 a_{2}b_{3} - a_{3}b_{1}\end{pmatrix}_{2_{3}} \\& \oplus \begin{pmatrix}
    a_{1}b_{3} +  a_{2}b_{1} + a_{3}b_{2}\\ -a_{1}b_{2} - a_{2}b_{3} -  a_{3}b_{1 }\end{pmatrix}_{2_{4}} \\
\end{align*}
 \begin{align*}
     \begin{pmatrix}
    a_1\\ a_2\\a_3\end{pmatrix}_{3_{1(2)}} \otimes   \begin{pmatrix}
    b_1\\b_2\\b_3\end{pmatrix}_{3_{2(1)}} = &\begin{pmatrix}
    a_{1}b_{1} + a_{2}b_{2} + a_{3}b_{3} \end{pmatrix}_{1_{2}}  \oplus \begin{pmatrix}
    a_{1}b_{1} + \omega^2 a_{2}b_{2} + \omega a_{3}b_{3 }\\ -\omega a_{1}b_{1} - \omega^2 a_{2}b_{2} -  a_{3}b_{3 }\end{pmatrix}_{2_{1}} \\&
    \oplus  \begin{pmatrix}
    a_{1}b_{2} + \omega^2 a_{2}b_{3} + \omega a_{3}b_{1 }\\-\omega a_{1}b_{3} - \omega^2 a_{2}b_{1} - a_{3}b_{2}\end{pmatrix}_{2_{2}}  \oplus  \begin{pmatrix}
    a_{1}b_{3} + \omega^2 a_{2}b_{1} + \omega a_{3}b_{2}\\ -\omega a_{1}b_{2} - \omega^2 a_{2}b_{3} - a_{3}b_{1}\end{pmatrix}_{2_{3}} \\& \oplus \begin{pmatrix}
    a_{1}b_{3} +  a_{2}b_{1} + a_{3}b_{2}\\ -a_{1}b_{2} - a_{2}b_{3} -  a_{3}b_{1 }\end{pmatrix}_{2_{4}} \\
\end{align*}
 \begin{align*}
    \begin{pmatrix}
    a_1\\ a_2\\a_3\end{pmatrix}_{3_{1(2)}} \otimes   \begin{pmatrix}
    b_1\\b_2\\b_3\end{pmatrix}_{3_{2(2)}} = \begin{pmatrix}
    a_{1}b_{1}\\a_{2}b_{2}\\a_{3}b_{3}\end{pmatrix}_{3_{2(1)}}  \oplus \begin{pmatrix}
    a_{3}b_{2} - a_{2}b_{3} \\a_{1}b_{3} - a_{3}b_{1} \\a_{2}b_{1} - a_{1}b_{2} \end{pmatrix}_{3_{1(1)}} \oplus  \begin{pmatrix}
    a_{3}b_{2} + a_{2}b_{3} \\a_{1}b_{3} + a_{3}b_{1} \\a_{2}b_{1} + a_{1}b_{2}\end{pmatrix}_{3_{2(1)}}   
\end{align*}
  \begin{align*}
     \begin{pmatrix}
    a_1\\ a_2\\a_3\end{pmatrix}_{3_{2(1)}} \otimes   \begin{pmatrix}
    b_1\\b_2\\b_3\end{pmatrix}_{3_{2(2)}} = &\begin{pmatrix}
    a_{1}b_{1} + a_{2}b_{2} + a_{3}b_{3} \end{pmatrix}_{1_{1}}  \oplus \begin{pmatrix}
    a_{1}b_{1} + \omega^2 a_{2}b_{2} + \omega a_{3}b_{3 }\\ \omega a_{1}b_{1} + \omega^2 a_{2}b_{2} +  a_{3}b_{3 }\end{pmatrix}_{2_{1}} \\&
    \oplus  \begin{pmatrix}
    a_{1}b_{2} + \omega^2 a_{2}b_{3} + \omega a_{3}b_{1 }\\ \omega a_{1}b_{3} + \omega^2 a_{2}b_{1} + a_{3}b_{2}\end{pmatrix}_{2_{2}}  \oplus  \begin{pmatrix}
    a_{1}b_{3} + \omega^2 a_{2}b_{1} + \omega a_{3}b_{2}\\ \omega a_{1}b_{2} + \omega^2 a_{2}b_{3} + a_{3}b_{1}\end{pmatrix}_{2_{3}} \\& \oplus \begin{pmatrix}
    a_{1}b_{3} +  a_{2}b_{1} + a_{3}b_{2}\\ a_{1}b_{2} + a_{2}b_{3} +  a_{3}b_{1 }\end{pmatrix}_{2_{4}} \\
\end{align*}

\subsection*{A.3 \quad Vacuum alignment}

We analyze the scalar potential to find out the vacuum alignment. The scalar potential
becomes rather simple in the $\Delta(54)$ symmetry. Especially, the supersymmetry is
important to see the vacuum alignment.
The invariant superpotential is given by
\begin{align*}
w &= {} \mu_1 \zeta^2 + \mu_2 \zeta^{\prime 6} +  \mu_3 \chi^2  \\
&+  \beta_1 \chi^\prime_1 \chi^\prime_2 + \beta^{\prime}_1 (\chi^{\prime 2 }_1 + \chi^{\prime 2 }_2)  \\
&+ \alpha_2 (\phi_{1}^{6} + \phi_{2}^{6} + \phi_{3}^{6}) + \alpha_2^{\prime} (\phi_{1}^2 \phi_{2}^2 \phi_{3}^2 )   \\
&+ \alpha_3 (\phi_{2}^{\prime6} + \phi_{2}^{\prime6} + \phi_{3}^{\prime6}) + \alpha_3^{\prime} (\phi_{1}^{\prime 2} \phi_{2}^{\prime 2}\phi_{3}^{\prime 2} )\\
&+ \alpha_4 (\phi_{2}^{\prime\prime6} + \phi_{2}^{\prime\prime6} + \phi_{3}^{\prime\prime6}) + \alpha_4^{\prime} (\phi_{1}^{\prime\prime 2} \phi_{2}^{\prime\prime 2}\phi_{3}^{\prime\prime 2} )\\
&+ \alpha_5 (\xi_{1}^{6} + \xi_{2}^{6} + \xi_{3}^{6}) + \alpha_5^{\prime} (\xi_{1}^2 \xi_{2}^2 \xi_{3}^2 )   \\
&+ \alpha_6 (\xi_{1}^{\prime6} + \xi_{2}^{\prime6} + \xi_{3}^{\prime6}) + \alpha_6^{\prime} (\xi_{1}^{\prime2} \xi_{2}^{\prime2} \xi_{3}^{\prime2} )   \\
\end{align*}
which leads to the scalar potential
 \begin{align*}
  V &= \lvert 2\mu_1 \zeta \rvert^2  +\lvert 6\mu_2 \zeta^{\prime 5} \rvert^2 + \lvert 2\mu_3 \chi \rvert^2 \\
&+  \lvert \beta_1 \chi^\prime_2 + 2 \beta^{\prime}_1 \chi^{\prime }_1 \rvert^2 
 +  \lvert \beta_1 \chi^\prime_1 + 2 \beta^{\prime}_1 \chi^{\prime }_2 \rvert^2 \\
 &+\lvert 6\alpha_2 \phi_{1}^5 + 2  \alpha_2^{\prime} \phi_{1}\phi_{2}^2\phi_{3}^2 \rvert^2 + \lvert 6\alpha_2 \phi_{2}^5 + 2  \alpha_2^{\prime} \phi_{2}\phi_{1}^2\phi_{3}^2 \rvert^2 +\lvert 6\alpha_2 \phi_{3}^5 + 2  \alpha_2^{\prime} \phi_{3}\phi_{1}^2\phi_{2}^2 \rvert^2  \\
  &+\lvert 6\alpha_3 \phi_{1}^{\prime5} + 2  \alpha_3^{\prime} \phi^{\prime}_{1}\phi_{2}^{\prime2}\phi_{3}^{\prime2} \rvert^2 + \lvert 6\alpha_3\phi_{2}^{\prime5} + 2  \alpha_3^{\prime} \phi^{\prime}_{2}\phi_{1}^{\prime2}\phi_{3}^{\prime2} \rvert^2 +\lvert 6\alpha_3 \phi_{3}^{\prime5} + 2  \alpha_3^{\prime} \phi^{\prime}_{3}\phi_{1}^{\prime2}\phi_{2}^{\prime2} \rvert^2 \\
  &+\lvert 6\alpha_4 \phi_{1}^{\prime\prime5} + 2  \alpha_4^{\prime} \phi^{\prime\prime}_{1}\phi_{2}^{\prime\prime2}\phi_{3}^{\prime\prime2} \rvert^2 + \lvert 6\alpha_4\phi_{2}^{\prime5} + 2  \alpha_4^{\prime} \phi^{\prime\prime}_{2}\phi_{1}^{\prime\prime2}\phi_{3}^{\prime\prime2} \rvert^2 +\lvert 6\alpha_4 \phi_{3}^{\prime5} + 2  \alpha_4^{\prime} \phi^{\prime\prime}_{3}\phi_{1}^{\prime\prime2}\phi_{2}^{\prime\prime2} \rvert^2 \\
 &+\lvert 6\alpha_5 \xi_{1}^5 + 2  \alpha_5^{\prime} \xi_{1}\xi_{2}^2\xi_{3}^2 \rvert^2 + \lvert 6\alpha_5 \xi_{2}^5 + 2  \alpha_5^{\prime} \xi_{2}\xi_{1}^2\xi_{3}^2 \rvert^2 +\lvert 6\alpha_5 \xi_{3}^5 + 2  \alpha_5^{\prime} \xi_{3}\xi_{1}^2\xi_{2}^2 \rvert^2  \\ 
 &+\lvert 6\alpha_6 \xi^{\prime5}_{1} + 2  \alpha^{\prime}_6 \xi^{\prime}_{1}\xi_{2}^{\prime2}\xi_{3}^{\prime2} \rvert^2 + \lvert 6\alpha_6 \xi_{2}^{\prime5} + 2  \alpha_6^{\prime} \xi^{\prime}_{2}\xi_{1}^{\prime2}\xi_{3}^{\prime2} \rvert^2 +\lvert 6\alpha_6 \xi_{3}^{\prime5} + 2  \alpha_6^{\prime} \xi^{\prime}_{3}\xi_{1}^{\prime2}\xi_{2}^{\prime2} \rvert^2  \\ 
 \end{align*}

The conditions of the potential minimum are written as
\begin{align*}
2\mu_1 \zeta &= 0 \\
 6\mu_2 \zeta^{\prime 5}  &= 0 \\
  2\mu_3 \chi  &= 0 \\
  \beta_1 \chi^\prime_2 + 2 \beta^{\prime}_1 \chi^{\prime }_1 &=0 ;& \qquad
  \beta_1 \chi^\prime_1 + 2 \beta^{\prime}_1 \chi^{\prime }_2 &=0 \\
  6\alpha_2 \phi_{1}^5 + 2 
 \alpha_2^{\prime} \phi_{1}\phi_{2}^2\phi_{3}^2  &= 0 ; &\qquad  6\alpha_2 \phi_{2}^5 + 2 
 \alpha_2^{\prime} \phi_{2}\phi_{1}^2\phi_{3}^2  &= 0 ; &\qquad  6\alpha_2 \phi_{3}^5 + 2 
 \alpha_2^{\prime} \phi_{3}\phi_{1}^2\phi_{2}^2  &= 0 \\
   6\alpha_3 \phi_{1}^{\prime5} + 2 
 \alpha_3^{\prime} \phi^{\prime}_{1}\phi_{2}^{\prime2}\phi_{3}^{\prime2}  &= 0 ; &\qquad  6\alpha_2 \phi_{2}^{\prime5} + 2 
 \alpha_3^{\prime} \phi^{\prime}_{2}\phi_{1}^{\prime2}\phi_{3}^{\prime2}  &= 0 ; &\qquad  6\alpha_3 \phi_{3}^{\prime5} + 2 
 \alpha_3^{\prime} \phi^{\prime}_{3}\phi_{1}^{\prime2}\phi_{2}^{\prime2}  &= 0 \\
   6\alpha_4 \phi_{1}^{\prime\prime5} + 2 
 \alpha_4^{\prime} \phi^{\prime\prime}_{1}\phi_{2}^{\prime\prime2}\phi_{3}^{\prime\prime2}  &= 0 ; &\qquad  6\alpha_4 \phi_{2}^{\prime\prime5} + 2 
 \alpha_4^{\prime} \phi^{\prime\prime}_{2}\phi_{1}^{\prime\prime2}\phi_{3}^{\prime\prime2}  &= 0 ; &\qquad  6\alpha_4 \phi_{3}^{\prime\prime5} + 2 
 \alpha_4^{\prime} \phi^{\prime\prime}_{3}\phi_{1}^{\prime\prime2}\phi_{2}^{\prime\prime2}  &= 0 \\
  6\alpha_5 \xi_{1}^5 + 2 
 \alpha_5^{\prime} \xi_{1}\xi_{2}^2\xi_{3}^2  &= 0 ; &\qquad  6\alpha_5 \xi_{2}^5 + 2 
 \alpha_5^{\prime} \xi_{2}\xi_{1}^2\xi_{3}^2  &= 0 ; &\qquad  6\alpha_5 \xi_{3}^5 + 2 
 \alpha_5^{\prime} \xi_{3}\xi_{1}^2\xi_{2}^2  &= 0 \\
 6\alpha_6 \xi_{1}^{\prime5} + 2 
 \alpha_6^{\prime} \xi^{\prime}_{1}\xi_{2}^{\prime2}\xi_{3}^{\prime2}  &= 0 ; &\qquad  6\alpha_6 \xi_{2}^{\prime5} + 2 
 \alpha_6^{\prime} \xi^{\prime}_{2}\xi_{1}^{\prime2}\xi_{3}^{\prime2}  &= 0 ; &\qquad  6\alpha_6 \xi_{3}^{\prime5} + 2 
 \alpha_6^{\prime} \xi^{\prime}_{3}\xi_{1}^{\prime2}\xi_{2}^{\prime2}  &= 0 \\
\end{align*}

A solution of these equations is
\begin{align*}
 \chi^\prime_1 =  \chi^\prime_2 &\qquad\text{with}\qquad \beta_1 + 2\beta^{\prime}_1 =0 \\
  \phi_{1} = \phi_{2} = \phi_{3} &\qquad\text{with}\qquad  3\alpha_2 + \alpha_2^{\prime} = 0\\
   \phi^{\prime}_{1} = \phi^{\prime}_{2} = \phi^{\prime}_{3} &\qquad\text{with}\qquad  3\alpha_3 + \alpha_3^{\prime} = 0\\
   \phi^{\prime\prime}_{1} = \phi^{\prime\prime}_{2} = \phi^{\prime\prime}_{3} &\qquad\text{with}\qquad  3\alpha_4 + \alpha_4^{\prime} = 0\\
 \xi_{1} = \xi_{2} = \xi_{3} &\qquad\text{with}\qquad  3\alpha_5 + \alpha_5^{\prime} = 0\\
 \xi^{\prime}_{1} = \xi^{\prime}_{2} = \xi^{\prime}_{3} &\qquad\text{with}\qquad  3\alpha_6 + \alpha_6^{\prime} = 0\\
 \end{align*}

 Therefore we can take Vacuum alignment as
 \begin{align*}
  \langle \chi^\prime \rangle& =(v_{\chi^\prime}, v_{\chi^\prime})&\\
   \langle \phi \rangle& =(v_{\phi}, v_{\phi}, v_{\phi})&\\
    \langle \phi^{\prime} \rangle& =(v^{\prime}_{\phi}, v^{\prime}_{\phi}, v^{\prime}_{\phi})&\\
     \langle \phi^{\prime\prime} \rangle& =(v^{\prime\prime}_{\phi}, v^{\prime\prime}_{\phi}, v^{\prime\prime}_{\phi})&\\
 \langle \xi \rangle& =(v_{\xi}, v_{\xi}, v_{\xi})&\\
  \langle \xi^{\prime} \rangle& =(v^{\prime}_{\xi}, v^{\prime}_{\xi}, v^{\prime}_{\xi})&\\
\end{align*}
\bibliographystyle{naturemag} 
\bibliography{references}

\end{document}